\documentclass[draft,journal,onecolumn]{IEEEtran}

\usepackage{cite}
\usepackage{graphicx}
\usepackage{amsmath}
\usepackage{amssymb}
\usepackage{amsthm}
\usepackage{amsfonts}
\usepackage{algorithmic}
\usepackage{algorithm}
\usepackage{mathrsfs}
\usepackage{cases}
\usepackage{mathtools}

\DeclareMathOperator{\LOC}{loc}
\DeclareMathOperator{\RANK}{rank}
\DeclareMathOperator{\SPAN}{span}
\DeclareMathOperator{\DIM}{dim}
\DeclareMathOperator{\GRANK}{rank_G}
\DeclareMathOperator{\ERANK}{rank_E}

\newtheorem{definition}{Definition}
\newtheorem{theorem}{Theorem}
\newtheorem{lemma}{Lemma}
\newtheorem{corollary}{Corollary}
\newtheorem{proposition}{Proposition}
\newtheorem{construction}{Construction}
\newtheorem{remark}{Remark}
\newtheorem{example}{Example}

\newcommand{\OVERSET}[2]{\overset{\mathclap{\scriptscriptstyle{#1}}}{#2}}

\begin{document}

\title{Locally Repairable Codes with\\
    Unequal Local Erasure Correction
}

\author{
    Geonu~Kim,~\IEEEmembership{Student~Member,~IEEE,} and~Jungwoo~Lee,~\IEEEmembership{Senior~Member,~IEEE}%
    \thanks{
        This research was supported in part by the Basic Science Research Program (NRF-2015R1A2A1A15052493)
        through NRF funded by MSIP, the Technology Innovation Program (10051928) funded by MOTIE,
        the Bio-Mimetic Robot Research Center funded by DAPA (UD130070ID), INMC, and BK21-plus.
    }%
    \thanks{
        The authors are with the Institute of New Media and Communications,
        Department of Electrical and Computer Engineering, Seoul National University, Seoul, 08826, Korea
        (e-mail: bdkim@wspl.snu.ac.kr; junglee@snu.ac.kr).
    }%
    \thanks{
        This paper was presented in part at the 2016 Allerton Conference on Communication, Control and Computing
        \cite{Kim16ALC}.
    }
}

\maketitle

\begin{abstract}
    When a node in a distributed storage system fails, it needs to be promptly repaired to maintain system integrity.
    While typical erasure codes can provide a significant storage advantage over replication,
    they suffer from poor repair efficiency.
    Locally repairable codes (LRCs) tackle this issue by reducing the number of nodes participating in the repair process
    (locality), at the cost of reduced minimum distance.
    In this paper, we study the tradeoff between locality and minimum distance of LRCs
    with local codes that have arbitrary distance requirements.
    Unlike existing methods where both the locality and the local distance requirements imposed on every node are identical,
    we allow the requirements to vary arbitrarily from node to node.
    Such a property can be an advantage for distributed storage systems with non-homogeneous characteristics.
    We present Singleton-type distance upper bounds and also provide an optimal code construction with respect to these bounds.
    In addition, the feasible rate region is characterized by dimension upper bounds that do not depend on the distance.
\end{abstract}

\IEEEpeerreviewmaketitle

\section{Introduction}

\subsection{Background} \label{subsect:BG}

The fundamental advantage of storing data in a distributed manner is that the risk of failure can be localized,
and a catastrophic loss at once of all the stored data can be avoided.
Furthermore, reliability of the system can be improved by using erasure codes on the user data
across different storage nodes in a \emph{distributed storage system} (DSS).
Although maximum distance separable (MDS) codes provide an optimal storage efficiency for a given amount of reliability,
they suffer from poor efficiency during \emph{repair} \cite{Sathiamoorthy13ICVLDB},
which is the recovery procedure for failed nodes.
Even if the number of failed storage nodes is below the erasure tolerance limit of the codes employed,
some (or all) of the failed nodes may have to be promptly repaired to maintain system integrity.
For a single node repair, while repetition codes only need access to another single node
which is just a replica of the failed node, MDS codes are an opposite extreme in that the number of required helper nodes
is as large as the dimension (number of information symbols) of the code.
Locally repairable codes (LRCs) try to minimize \emph{locality},
which is the number of nodes that are accessed during repair,
for given code parameters such as length, dimension, and minimum distance.
The tradeoff between locality and other parameters has been studied extensively
since the discovery of the Singleton-type bound in \cite{Gopalan12TIT}.

A natural extension to the conventional locality is the \emph{$(r,\delta)$-locality} \cite{Prakash12ISIT,Kamath14TIT},
where more flexible repair options are provided by generalizing the constraint on the minimum distance of local codes
to at least $\delta$ instead of 2 (single parity checks).
Such flexibility is beneficial to modern large-scale DSSs where multiple node failures have become more common.
For example, in conventional optimal LRCs \cite{Gopalan12TIT,Papailiopoulos14TIT,Tamo14TIT} with locality $r$,
if another node included in the local repair group of a failed node simultaneously fails,
repair from $r$ nodes is no longer valid,
and a large number of nodes have to be accessed to perform ordinary erasure correction.
On the other hand, $(r,\delta)$-LRCs can still perform repair from $r$ nodes
even if $\delta-1$ nodes in a local repair group simultaneously fail.

Recently, there has been interest in the case where locality
is specified differently for different nodes \cite{Kadhe16ISIT,Kadhe16ARX,Zeh16ISIT,Zeh16ARX}.
Such situations may occur, for example, when the underlying storage network is not homogeneous.
It would also be beneficial in the scenarios where \emph{hot data} symbols require faster repair or reduced download latency
\cite{Kadhe16ISIT}.
In \cite{Kadhe16ISIT,Zeh16ISIT}, relevant Singleton-type bounds have been found
and some optimal code constructions are also given, which show the tightness of the bounds.

\subsection{Contributions and Organization}

In this paper, we study the tradeoff between locality and minimum distance for $(r,\delta)$-LRCs,
where the locality parameter $r$ and the local distance parameter $\delta$ are not necessarily the same for each node.
Our main contribution is different from previous work on unequal locality \cite{Kadhe16ISIT,Zeh16ISIT} in two ways.
First, we extend the results on conventional \emph{$r$-locality} to \emph{$(r,\delta)$-locality}.
Specifically, our new Singleton-type bound based on the notion of \emph{layered $(r,\delta)$-locality}
(Proposition \ref{prop:LL-distbound}), which is a generalization of the notion of \emph{locality profile} \cite{Kadhe16ISIT},
includes the bound in \cite{Kadhe16ISIT} as a special case. 
Second, and more importantly, we present a bound (Theorem \ref{thm:UL-distbound})
whose expression is directly based on the \emph{unequal $(r,\delta)$-locality requirement}
without the undesirable \emph{layered} constraint,
given that the unequal $(r,\delta)$ parameters satisfy a natural ordering condition.
Moreover, this bound is shown to be tight in the sense that
non-trivial codes (Construction \ref{cnstrct}) that achieve the equality in the bound exist (Theorem \ref{thm:optOL}).
We also characterize the feasible rate region by an upper bound on the code dimension,
which does not depend on the minimum distance
(Proposition \ref{prop:LL-dimbound} and Theorem \ref{thm:UL-dimbound}).

The rest of this paper is organized as follows.
In Section \ref{sect:preliminaries}, we review some important preliminaries.
Section \ref{sect:UL&LL} describes the motivation of our work
and provides formal definitions for both layered and unequal $(r,\delta)$-locality.
Our Singleton-type bounds based on layered and unequal $(r,\delta)$-locality
are provided respectively in Section \ref{sect:LLBND} and \ref{sect:ULBND},
together with the corresponding dimension upper bounds.
Section \ref{sect:construction} shows a code construction scheme
that is optimal for the bounds in Section \ref{sect:ULBND}.
In Section \ref{sect:etc}, we provide some further results, including the optimality of the code construction
in Section \ref{sect:construction} in terms of the bound in Section \ref{sect:LLBND},
and a further tightened bound in the two different ordered $(r,\delta)$-locality case.
Finally, the concluding remarks are drawn in Section \ref{sect:conclusion}.

\section{Preliminaries} \label{sect:preliminaries}

\subsection{Notation}

We use the following notation.

\begin{enumerate}
    \item For an integer $i$, $ [i] = \{ 1, \ldots, i \} $.
    \item A vector of length $n$ is denoted by $ \mathbf{v} = (v_1,\ldots,v_n) $.
    \item A matrix of size $ k \times n $ is denoted by $ G = (g_{i,j})_{ i \in [k], j \in [n]} $.
    \item For sets $\mathcal{A}$ and $\mathcal{B}$, $ \mathcal{A} \sqcup \mathcal{B} $ denotes the disjoint union, i.e.,
        $ \mathcal{A} \cup \mathcal{B} $ with further implication that $ \mathcal{A} \cap \mathcal{B} = \emptyset $.
    \item For a symbol index set $ \mathcal{T} \subset [n] $ of a code $\mathscr{C}$ of length $n$,
        $\mathscr{C}\rvert_{\mathcal{T}}$ denotes the punctured code with support $\mathcal{T}$,
        and $G\rvert_{\mathcal{T}}$ is the corresponding generator matrix. Furthermore, we define
        $ \GRANK(\mathcal{T}) = \RANK(G\rvert_{\mathcal{T}}) $.
    \item For a symbol index set $ \mathcal{T} \subset [n] $ of a linear $[n,k]$ code $ \mathscr{C} $
        constructed via polynomial evaluation on an extension field $\mathbb{F}_{q^t}$,
        $\ERANK(\mathcal{T})$ denotes the rank of the evaluation points corresponding to $\mathscr{C}\rvert_{\mathcal{T}}$
        over the base field $\mathbb{F}_q$.
\end{enumerate}

\subsection{Minimum Distance}

The minimum distance of linear codes is well known to be characterized by the following lemma
\cite[Lem. A.1]{Kamath14TIT}, which is the basis of our minimum distance bounds.

\begin{lemma} \label{lem:dist-rank}
    For a symbol index set $ \mathcal{T} \subset [n]$ of a linear $[n,k,d]$ code such that $ \GRANK(\mathcal{T}) \leq k - 1 $, 
    we have 
    \begin{equation*}
        d \leq n - \lvert \mathcal{T} \rvert \text{,}
    \end{equation*}
    with equality if $\mathcal{T}$ is of largest cardinality.
\end{lemma}

Below, we state a lemma (see also the proof of \cite[Thm. 1.1]{Kuijper14ARX}) based on Lemma \ref{lem:dist-rank}
that turns out to be more useful.
Note that Lemma \ref{lem:dist-red} can not be derived by simply substituting $\lvert \mathcal{T} \rvert$
into Lemma \ref{lem:dist-rank}.

\begin{lemma} \label{lem:dist-red}
    For a symbol index set $ \mathcal{T} \subset [n]$ of a linear $[n,k,d]$ code such that $ \GRANK(\mathcal{T}) \leq k - 1 $,
    let $\gamma$ be the number of redundant symbols indexed by $\mathcal{T}$,
    i.e., $ \gamma = \lvert \mathcal{T} \rvert - \GRANK(\mathcal{T}) $.
    We have
    \begin{equation*}
        d \leq n - k + 1 - \gamma \text{.}
    \end{equation*}
\end{lemma} 

\begin{IEEEproof}
    Clearly, the set $\mathcal{T}$ can be enlarged to a set $\mathcal{T}'$ such that $ \GRANK(\mathcal{T}') = k - 1 $.
    Make another set $\mathcal{T}''$ by removing $\gamma$ redundant symbols from $\mathcal{T}'$.
    Note that $ \lvert \mathcal{T}'' \rvert \geq k - 1 $ since $ \GRANK(\mathcal{T}'') = k - 1 $.
    By applying Lemma \ref{lem:dist-rank} to the set $\mathcal{T}'$, we have
    \begin{align*}
        d & \leq n - \lvert \mathcal{T}' \rvert = n - \lvert \mathcal{T}'' \rvert - \gamma\\
        & \leq n - k + 1 - \gamma \text{.}
    \end{align*}
\end{IEEEproof}

As an immediate corollary to Lemma \ref{lem:dist-rank}, we also get the following lemma,
which is used when showing the optimal distance property of our code construction.

\begin{lemma} \label{lem:dist-rank:cor}
    Given any linear $[n,k,d]$ code, for every symbol index set $ \mathcal{T} \subset [n] $ of cardinality 
    $ \lvert \mathcal{T} \rvert = \tau $ with $ \GRANK(\mathcal{T}) = k $, we have
    \begin{equation*}
        d \geq n - \tau + 1 \text{.}
    \end{equation*}
\end{lemma}

\begin{remark} \label{rem:rank}
    In Lemma \ref{lem:dist-rank}, \ref{lem:dist-red}, and \ref{lem:dist-rank:cor},
    erasure correction is possible from $\mathcal{T}$ if and only if $ \GRANK(\mathcal{T}) = k $.
    Equivalently, erasure correction is not possible from $\mathcal{T}$ if and only if $ \GRANK(\mathcal{T}) \leq k - 1 $.
\end{remark}

\subsection{$(r,\delta)$-Locality}

A linear $[n,k,d]$ code $\mathscr{C}$ is said to have \emph{locality} $r$ (or $r$-locality)
if every symbol of $\mathscr{C}$ can be recovered with a linear combination of at most $r$ other symbols \cite{Gopalan12TIT},
i.e.,
\begin{equation*}
    \LOC(i) \leq r \text{,}
\end{equation*}
for every symbol index $ i \in [n] $,
where $\LOC(i)$ denotes the \emph{smallest} number of other symbols that allow the recovery of the $i$th symbol.
An equivalent description is that for each symbol index $ i \in [n] $, there exists a punctured code of $\mathscr{C}$ with
support containing $i$, of length at most $ r + 1 $ and distance of at least $2$.
We call such codes $r$-LRCs.
It has been shown in \cite{Gopalan12TIT} that the minimum Hamming distance $d$ of an $[n,k,d]$ $r$-LRC is upper bounded by
\begin{equation}
    d \leq n - k + 2 - \left\lceil \frac{k}{r} \right\rceil \text{,} \label{eq:GBND}
\end{equation}
which reduces to the well-known \emph{Singleton bound} if $ r \geq k $.
Various optimal code constructions achieving the equality in the minimum distance bound have been reported in the literature
\cite{Gopalan12TIT,Tamo13ISIT,Silberstein13ISIT,Song14JSAC,Tamo14TIT,Papailiopoulos14TIT,Goparaju14ISIT,Tamo15ISIT,Hao16ISIT}.
It has also been shown in \cite{Tamo14TIT} that the rate of an $[n,k]$ $r$-LRC is upper bounded by
\begin{equation*}
    \frac{k}{n} \leq \frac{r}{r+1} \text{,}
\end{equation*}
which can also be written as a dimension upper bound that does not depend on the minimum distance, i.e.,
\begin{equation*}
    k \leq \frac{n}{r+1} \cdot r \text{.}
\end{equation*}

The notion of $r$-locality can be naturally extended to $(r,\delta)$-locality \cite{Prakash12ISIT}
to address the situation with multiple (local) node failures.
Note that $r$-locality corresponds to $(r,\delta=2)$-locality.

\begin{definition}[$(r,\delta)$-locality] \label{def:DLOC}
    For integers $ r \geq 1 $ and $ \delta \geq 2 $, a symbol with index $ i \in [n] $
    of a linear $[n,k]$ code $\mathscr{C}$ is said to have $(r,\delta)$-locality,
    if there exists a punctured code of $\mathscr{C}$ with support containing $i$,
    of length at most $ r + \delta - 1 $ and distance of at least $\delta$,
    i.e., there exists a symbol index set $ \mathcal{S}_i \subset [n] $ such that
    \begin{itemize}
        \item $ i \in \mathcal{S}_i $,
        \item $ \lvert \mathcal{S}_i \rvert \leq r + \delta - 1 $,
        \item $ d( \mathscr{C}\rvert_{\mathcal{S}_i} ) \geq \delta $.
    \end{itemize}
\end{definition}

Furthermore, $\mathscr{C}$ in the definition above is said to have $(r,\delta)$-locality
if every symbol has $(r,\delta)$-locality, and is also called an $(r,\delta)$-LRC.
We have the following remark \cite{Kamath14TIT}.

\begin{remark} \label{rem:LCSB}
    By applying the Singleton bound to $\mathscr{C}\rvert_{\mathcal{S}_i}$ in Definition \ref{def:DLOC},
    we get $ \GRANK(\mathcal{S}_i) \leq r $.
\end{remark}

It has been shown in \cite{Prakash12ISIT,Kamath14TIT} that the minimum distance of an $(r,\delta)$-LRC is upper bounded by
\begin{equation}
    d \leq n - k + 1 - \left( \left\lceil \frac{k}{r} \right\rceil - 1 \right) ( \delta - 1 ) \text{.} \label{eq:PBND}
\end{equation}
There are also several optimal code constructions in the literature
\cite{Prakash12ISIT,Kamath14TIT,Silberstein13ISIT,Tamo13ISIT,Tamo14TIT,Song14JSAC,Ernvall16TIT,Poellaenen16ISIT,Chen16ARX}
that achieve the equality in \eqref{eq:PBND}.
The rate upper bound for $(r,\delta)$-LRCs has been shown in \cite{Song14JSAC,Song15ARX} to be
\begin{equation*}
    \frac{k}{n} \leq \frac{r}{ r + \delta + 1 } \text{,}
\end{equation*}
which can also be expressed as an upper bound on the dimension as 
\begin{equation}
    k \leq \frac{n}{ r + \delta + 1 } \cdot r \text{.} \label{eq:EL-dimbound}
\end{equation}

\subsection{$r$-LRC with Unequal Locality}

Consider DSSs that require different $r$-localities for different nodes.
In other words, the upper limit for $\LOC(\cdot)$ varies from symbol to symbol for the code employed in the DSS.
Among the different locality requirements, let us denote the minimum locality requirement by $r_{min}$.
We can employ $r_{min}$-LRCs where $ \LOC(\cdot) \leq r_{min} $ for every symbol.
However, this clearly tends to be an \emph{over-design},
and one may expect improved minimum distance by taking advantage of the looser locality constraints.

To be more specific, suppose that for a linear $[n,k,d]$ code and
symbol index sets $ \mathcal{N}_1,\mathcal{N}_2 \subsetneq [n] $ such that $ \mathcal{N}_1 \sqcup \mathcal{N}_2 = [n] $,
we require $ \LOC(i_1) \leq r_1 $ and $ \LOC(i_2) \leq r_2 $,
where $ i_1 \in \mathcal{N}_1 $, $ i_2 \in \mathcal{N}_2 $ and $ r_1 < r_2 $.
Clearly $r_1$-LRCs satisfy this requirement and by optimal code constructions with respect to \eqref{eq:GBND}, we may achieve
\begin{equation*}
    d = d_1 \triangleq n - k + 2 - \left\lceil \frac{k}{r_1} \right\rceil \text{.}
\end{equation*}
On the other hand, any code that fulfills the DSS requirement of unequal locality is clearly an $r_2$-LRC.
Therefore, again by \eqref{eq:GBND}, we have
\begin{equation*}
    d \leq d_2 \triangleq n - k + 2 - \left\lceil \frac{k}{r_2} \right\rceil \text{.}
\end{equation*}
Note that $ d_1 \leq d_2 $ where equality does not hold in general.
Now the question is as follows: can we construct codes achieving a distance larger than $d_1$, and at the same time,
do we have a distance bound tighter than $d_2$?

\begin{example} \label{ex:UL:req}
    Suppose that the required DSS code parameters are $ r_1 = 2 $, $ r_2 = 5 $, $ n_1 = \lvert \mathcal{N}_1 \rvert = 6 $,
    $ n_2 = \lvert \mathcal{N}_2 \rvert = 24 $, and $ k = 19 $.
    $[n=n_1+n_2=30,k=19,d]$ $(r=r_1=2)$-LRCs clearly satisfy the DSS requirement,
    and there exist such LRCs that are distance optimal with respect to the bound by \eqref{eq:GBND},
    so that $ d = d_1 \triangleq 3 $.
    However, since the looser locality constraint of $ r_2 = 5 $ is not exploited,
    it is expected that there exist codes satisfying the DSS requirement with minimum distance $d$ larger than $ d_1 = 3 $.
    On the other hand, noting that the codes under specification are also $(r=r_2=5)$-LRCs,
    \eqref{eq:GBND} gives a trivial minimum distance upper bound of $ d \leq d_2 \triangleq 9 $.
    Clearly, this bound is expected to be loose, since the DSS constraint is stronger than the $(r=r_2=5)$-LRC constraint.
\end{example}

Recent works \cite{Kadhe16ISIT,Zeh16ISIT} have provided answers to the question above, but with some restrictions
that differ between \cite{Kadhe16ISIT} and \cite{Zeh16ISIT}.

In \cite{Kadhe16ISIT}, it is assumed that the locality for each symbol is specified in a minimum sense.
In other words, the constraint by the locality parameter $r$ specified on the $i$th symbol requires that $ \LOC(i) = r $,
which is equivalent to saying that the $i$th symbol has $r$-locality, i.e., $ \LOC(i) \leq r $,
but not $r'$-locality such that $ r' = r - 1 $, i.e., $ \LOC(i) \nleq r - 1 $.
The notion of \emph{locality profile} captures this minimum locality specification for every symbol of the code,
while the conventional specification as an upper bound on $\LOC(\cdot)$ is called the \emph{locality requirement}.
In particular, the locality profile of a linear $[n,k,d]$ code $\mathscr{C}$ is defined as the vector
$ \mathbf{r} = ( r_1, \ldots, r_n ) $, where $ r_i = \LOC(i) $, $ i \in [n] $.
The locality profile can also be specified as another vector $ \mathbf{n} = ( n_1, \ldots, n_{r^*} ) $,
where $ r^* = \max\{ r_1, \ldots, r_n \} $ and $n_j$ is the number of symbols such that $\LOC(\cdot)$ equals $j$,
for $ j \in [r^*] $.
The minimum Hamming distance of $\mathscr{C}$ is shown to be upper bounded by
\begin{equation}
    d \leq n - k + 2 - \sum_{j=1}^{r-1} \left\lceil \frac{n_j}{ j + 1 } \right\rceil
        - \left\lceil \frac{ k - \sum_{j=1}^{r-1} ( n_j - \left\lceil \frac{n_j}{ j + 1 } \right\rceil ) }{r} \right\rceil
        \text{,} \label{eq:Kadhe}
\end{equation}
where
\begin{equation*}
    r = \min{\{ j \in [r^*] \mid \sum_{j'=1}^j ( n_{j'} - \left\lceil \frac{n_{j'}}{ {j'} + 1 } \right\rceil ) \geq k \}}
        \text{.}
\end{equation*}
Furthermore, this bound is demonstrated to be tight by an optimal code construction
based on Gabidulin outer codes \cite{Silberstein13ISIT}, achieving the equality in the bound for some parameter regime.

\begin{example} \label{ex:UL:prf}
    Consider linear $[n=30,k=19,d]$ codes with locality profile $\mathbf{n}=(0,6,0,0,24)$.
    Since there exist distance optimal $[n=30,k=19,d=d_1\triangleq3]$ $(r=r_1=2)$-LRCs
    with locality profile $\mathbf{n}=(0,30)$, which is a stronger restriction than the required locality profile,
    we expect the existence of relevant codes with minimum distance $d$ larger than $ d_1 = 3 $.
    On the other hand, as in Example \ref{ex:UL:req}, we have $ d \leq d_2 \triangleq 9 $,
    since the considered codes are also $(r=r_2=5)$-LRCs.
    The bound by \eqref{eq:Kadhe} shows that we have in fact $ d \leq d_{UB} \triangleq 8 $.
    Furthermore, the construction based on Gabidulin outer codes gives relevant codes of optimal minimum distance
    $ d = d_{UB} = 8 $.
\end{example}

Note that the problem setting of Example \ref{ex:UL:prf} is more restrictive than that in Example \ref{ex:UL:req},
and therefore does not provide an answer for the original problem of Example \ref{ex:UL:req}.
In particular, going back to the original unequal DSS locality requirements of Example \ref{ex:UL:req},
we can not simply eliminate the possibility that codes of minimum distance larger than $ d = d_{UB} = 8 $
and locality profile other than $(0,6,0,0,24)$ exist.

The work by \cite{Zeh16ISIT} has investigated the same problem of unequal locality requirement,
but with a kind of \emph{disjointness}%
\footnote{
    The repair group of a symbol having $r_j$-locality consists of symbols in $\mathcal{N}_j$ only,
    where $\mathcal{N}_j$ is the index set of symbols having $r_j$-locality.
}
constraint.
It turns out that the disjointness restriction plays a similar role as the restriction by the locality profile
in the derivation of the minimum distance upper bound, resulting in an expression similar to \eqref{eq:Kadhe}.

\subsection{Gabidulin Codes}

Our optimal code construction is an extension of the LRC construction based on Gabidulin codes
\cite{Silberstein13ISIT,Kadhe16ISIT}.
We thus give a brief introduction on Gabidulin codes, including some relevant properties.

Due to the vector space structure of extension fields, an element in $\mathbb{F}_{q^t}$ can be
equivalently expressed as a vector of length $t$ over the base field $\mathbb{F}_q$, i.e., $\mathbb{F}_q^t$.
Consequently, a vector $ \mathbf{v} \in \mathbb{F}_{q^t}^n $
can be represented as a matrix $ V \in \mathbb{F}_q^{t \times n} $
where each column vector of the matrix $V$ corresponds to the vector representation of an element in vector $\mathbf{v}$.
The rank of the vector $\mathbf{v}$ is defined as $ \RANK(\mathbf{v}) = \RANK(V) $.
Furthermore, a \emph{metric} called \emph{rank distance} can be defined
for two vectors $ \mathbf{u},\mathbf{v} \in \mathbb{F}_{q^t}^n $ as
\begin{equation*}
    d_R( \mathbf{u},\mathbf{v} ) \triangleq \RANK( \mathbf{u} - \mathbf{v} ) = \RANK( U - V ) \text{.}
\end{equation*}
It is easy to see that the rank distance is upper bounded by the Hamming distance, i.e.,
$ d_R( \mathbf{u}, \mathbf{v} ) \leq d_H(\mathbf{u}, \mathbf{v}) $.
Therefore the minimum rank distance of a linear $[n,k]_{q^t}$ code is also upper bounded by the Singleton bound,
and the codes achieving this bound are called maximum rank distance (MRD) codes. Clearly, MRD codes also have the MDS property.

Gabidulin codes \cite{Gabidulin85} are an important class of codes with the MRD property.
Similar to Reed-Solomon and other algebraic codes, Gabidulin codes are constructed via polynomial evaluation.
However, both the data polynomials and the evaluation points are different.
In particular, an $[n,k]_{q^t}$ Gabidulin code ($ t \geq n $) is constructed by encoding a message vector
$ \mathbf{a} = ( a_1, \ldots, a_k ) \in \mathbb{F}_{q^t}^k $ according to the following two steps.
\begin{enumerate}
    \item Construct a data polynomial $ f(x) = \sum_{i=1}^k a_i x^{q^{i-1}} $.
    \item Obtain a codeword by evaluating $f(x)$ at $n$ points $ \{ x_1, \ldots, x_n \} \subset \mathbb{F}_{q^t} $
        (or $\mathbb{F}_q^t$) that are linearly independent over $\mathbb{F}_q$, i.e.,
        $ \mathbf{c} = ( f(x_1), \ldots, f(x_n) ) \in \mathbb{F}_{q^t}^n $
        with $ \RANK(\{ x_1, \ldots, x_n \}) = n $.
\end{enumerate}
The data polynomial $f(x)$ belongs to a special class of polynomials
called \emph{linearized polynomials} \cite{Macwilliams77Book}.
The evaluation of a linearized polynomial over $\mathbb{F}_{q^t}$ is an $\mathbb{F}_q$-linear transformation.
In other words, for any $ a,b \in \mathbb{F}_q $ and $ x,y \in \mathbb{F}_{q^t} $, the following holds.
\begin{equation}
    f( ax + by ) = a f(x) + b f(y) \text{.} \label{eq:fqlin}
\end{equation}
The rank distance of any Gabidulin codeword $\mathbf{c}$ can be shown to meet the Singleton bound by noting that
\begin{equation*}
    \RANK(\mathbf{c}) = \DIM(\SPAN(\{ f(x_1), \ldots, f(x_n) \})) \OVERSET{\eqref{eq:fqlin}}{=}
        \DIM(f(\SPAN(\{ x_1, \ldots, x_n \}))) \OVERSET{(a)}{\geq} n - ( k - 1 ) \text{,}
\end{equation*}
where (a) is due to the \emph{rank-nullity theorem}
and the fact that the nullity of $f(\cdot)$ is at most the $q$-degree of $f(\cdot)$, i.e., $ k - 1 $.

Although it is sufficient to claim Gabidulin codes to be MDS by their MRD property,
a more insightful derivation can be obtained by showing the MDS property directly with the 
analysis of their erasure correction capability.
Specifically, the polynomial $f(\cdot)$, and therefore the underlying message vector $\mathbf{a}$,
can be recovered from evaluations on any $k$ points $\{ f(y_1), \ldots, f(y_k) \}$
that are linearly independent (over $\mathbb{F}_q$), i.e., $ \RANK(\{ y_1, \ldots, y_k \}) = k $.
This argument is true since the use of the $\mathbb{F}_q$-linearity in \eqref{eq:fqlin}
makes it possible to obtain evaluations at $q^k$ different points,
from which the polynomial $f(\cdot)$ of degree $q^{k-1}$ can be interpolated.
Therefore, erasure correction is possible from arbitrary $k$ symbols of the codeword.

More importantly, note that the evaluation points may differ from the original ones used in the codeword construction.
This turns out to be the case for our optimal code construction,
where we apply MDS encoding on chunks of a Gabidulin codeword to equip the code with the desired locality property.
To analyze the possibility of erasure correction (or decodability) of an erasure pattern of the code,
all we need to do is to figure out whether the \emph{remaining rank}, which refers to the rank of the evaluation points
corresponding to the remaining symbols of the erasure pattern, is at least $k$.
In other words, the number of tolerable \emph{rank erasures} is at most $ n - k $.
The following lemma, which is a special case of \cite[Lem. 9]{Rawat14TIT},
will be used several times in analyzing the distance of our optimal code construction.

\begin{lemma} \label{lem:ERANK}
    For a vector $\mathbf{u}$ of length $k$ with elements being evaluations of a linearized polynomial $f(\cdot)$
    over $\mathbb{F}_{q^t}$, such that the evaluation points are linearly independent over $\mathbb{F}_q$,
    let $\mathbf{v}$ be the vector obtained by encoding $\mathbf{u}$ with an $[n,k]_q$ MDS code.
    Then any $s$ symbols of the codeword $\mathbf{v}$ correspond to the evaluations of $f(\cdot)$ at $s$ points
    lying in the subspace spanned by the original $k$ evaluation points (of $\mathbf{u}$) with rank $\min(s,k)$,
    i.e., for an arbitrary set $ \mathcal{T} \subset [n] $ such that $ \lvert \mathcal{T} \rvert = s $, we have
    \begin{equation*}
        \ERANK(\mathcal{T}) = \min(s,k) \text{.}
    \end{equation*}
\end{lemma}

\section{Unequal Locality and Layered Locality} \label{sect:UL&LL}

Both $(r,\delta)$-locality \cite{Prakash12ISIT} and unequal $r$-locality%
\footnote{
    We usually simply use the term locality to denote $(r,\delta)$-locality.
    The conventional case of $ \delta = 2 $ will be explicitly referred as $r$-locality.
}
\cite{Kadhe16ISIT,Zeh16ISIT} are very useful concepts
providing flexibility in the repair of multiple symbols and hot data symbols.
It is therefore natural to combine and benefit from both of the ideas.%
\footnote{
    The independent work by \cite{Chen17ISIT} also studies the same problem,
    but under the additional constraint of \emph{disjointness} as in \cite{Zeh16ISIT}.
}

\begin{definition}[Unequal locality] \label{def:UL}
    Let $ [n] = \bigsqcup_{j=1}^{s^*} \mathcal{N}_j $ and $ \lvert\mathcal{N}_j\rvert = n_j $, $ j \in [s^*] $.
    A linear $[n,k]$ code $\mathscr{C}$ is said to have unequal locality with parameters $\{(n_j,r_j,\delta_j)\}_{j\in[s^*]}$,
    where $ (r_j,\delta_j) \neq (r_{j'},\delta_{j'}) $ for $ j \neq j' $, $ j, j' \in [s^*] $,
    if every symbol with index $ i \in \mathcal{N}_j $, $ j \in [s^*] $,
    has $(r_j,\delta_j)$-locality.
    Furthermore, define
    \begin{itemize}
        \item integers $p_j$, $q_j$ such that $ n_j = p_j( r_j + \delta_j - 1 ) + q_j $
            and $ 0 \leq q_j \leq r_j + \delta_j - 2 $,
        \item $ \displaystyle m_j \triangleq \frac{n_j}{ r_j + \delta_j - 1 } = p_j + \frac{q_j}{ r_j + \delta_j - 1 } $,
        \item $ \displaystyle k_j \triangleq \begin{cases}
                \lfloor m_j \rfloor r_j & \text{if $ 0 \leq q_j \leq \delta_j - 2 $,} \\
                n_j - \lceil m_j \rceil ( \delta_j - 1 ) & \text{if $ \delta_j - 1 \leq q_j \leq r_j + \delta_j - 2 $.}
            \end{cases} $
    \end{itemize}
\end{definition}

More useful results are obtained in the special case where the parameters $r_j$ and $\delta_j$ follow
the two ordering conditions below. (See also \cite[Def. 4]{Chen17ISIT}.)

\begin{definition}[Ordered-$(r,\delta)$] \label{def:ORDCON}
    The unequal $\{(r_j,\delta_j)\}_{j\in[s^*]}$-locality parameters are said to satisfy the ordered $(r,\delta)$ condition if
    \begin{itemize}
        \item $ r_1 \leq r_2 \leq \cdots \leq r_{s^*} $,
        \item $ \delta_1 \geq \delta_2 \geq \cdots \geq \delta_{s^*} $.
    \end{itemize}
\end{definition}

Clearly, either condition alone can be assumed without loss of generality.
The ordered $(r,\delta)$ condition is therefore always satisfied if $ \delta_1 = \delta_2 = \cdots \delta_{s^*} $.
Note that the special case of $ \delta_1 = \delta_2 = \cdots \delta_{s^*} = 2 $ results in unequal $r$-locality.

Moreover, the ordered $(r,\delta)$ condition is natural in that
both smaller locality $r_j$ and larger local distance $\delta_j$ imply higher priority on the corresponding symbols.
In particular, symbols of smaller locality $r_j$ and larger local distance $\delta_j$ can be repaired more quickly and
are more robust to node failures.
These properties are well suited for \emph{hot data} symbols,
and therefore, codes having unequal locality with ordered-$(r,\delta)$ are of special interest.

The following example describes our main problem to be solved.

\begin{example} \label{ex:main:UL}
    Consider linear $[n=30,k=13]$ codes having $\{(n_1=6,r_1=3,\delta_1=4),(n_2=24,r_2=5,\delta_2=2)\}$-locality.
    Since $(r_1=3,\delta_1=4)$-locality implies $(r_2=5,\delta_2=2)$-locality,
    $[n=30,k=13]$ $(r=r_1=3,\delta=\delta_1=4)$-LRCs are among such codes.
    Furthermore, there exist optimal $[n=30,k=13,d=d_1\triangleq6]$ $(r=r_1=3,\delta=\delta_1=4)$-LRCs
    with respect to the bound by \eqref{eq:PBND}.
    However, since the weaker locality constraint of $(r_2=5,\delta_2=2)$ is not exploited,
    it is expected that there exist codes with minimum distance $d$ larger than $ d_1 = 6 $.
    On the other hand, by observing that the codes under consideration are also $(r=r_2=5,\delta=\delta_2=2)$-LRCs,
    we readily have the trivial minimum distance upper bound of $ d \leq d_2 \triangleq 16 $, again by \eqref{eq:PBND}.
    However, this bound is expected to be loose,
    since the considered constraint is stronger than the $(r=r_2=5,\delta=\delta_2=2)$-LRC constraint.
\end{example}

Directly studying \emph{unequal locality} even in its simpler form with $ \delta_j = 2 $, $ j \in [s^*] $,
appears to be intractable and some further restricting conditions have been used in \cite{Kadhe16ISIT,Zeh16ISIT}.
We also make use of the notion of \emph{layered locality},
which can be seen as a generalization of the \emph{locality profile} \cite{Kadhe16ISIT}, as an intermediate step.%
\footnote{
    The notion of unequal locality (Definition \ref{def:UL}) can be seen as
    a generalization of the \emph{locality requirement} \cite{Kadhe16ISIT}.
}

\begin{definition}[Layered locality] \label{def:LL}
    Let $ [n] = \bigsqcup_{j=1}^{s^*} \mathcal{N}_j $ and $ \lvert\mathcal{N}_j\rvert = n_j $, $ j \in [s^*] $.
    A linear $[n,k]$ code $\mathscr{C}$ is said to have layered locality with parameters $\{(n_j,r_j,\delta_j)\}_{j\in[s^*]}$,
    where $ (r_j,\delta_j) \neq (r_{j'},\delta_{j'}) $ for $ j \neq j' $, $ j, j' \in [s^*] $,
    if every symbol with index $ i \in \mathcal{N}_j $, $ j \in [s^*] $,
    has $(r_j,\delta_j)$-locality but not $(r_{j'},\delta_{j'})$-locality, $ j' \in [j-1] $.
    Parameters $p_j$, $q_j$, $m_j$ and $k_j$ are defined in the same way as for the unequal locality (Definition \ref{def:UL}).
    Furthermore, define the following incremental rank parameters, $ j \in [s^*] $.
    \begin{equation*}
        \xi_j \triangleq \GRANK(\bigsqcup_{j'=1}^j\mathcal{N}_{j'}) - \GRANK(\bigsqcup_{j'=1}^{j-1}\mathcal{N}_{j'}) \text{.}
    \end{equation*}
\end{definition}

It is possible that codes of certain layered locality do not exist.
Consider, for example, codes having layered $\{(n_1,r_1,\delta_1),(n_2,r_2,\delta_2)\}$-locality
where $ r_1 \geq r_2 $ and $ \delta_1 \leq \delta_2 $.
Since $(r_2,\delta_2)$-locality implies $(r_1,\delta_1)$-locality, it must be true that $ n_2 = 0 $.
Having $ n_2 > 0 $ for the layered $\{(n_1,r_1,\delta_1),(n_2,r_2,\delta_2)\}$-locality is therefore contradictory
or \emph{improper}.
Note that, the ordered $(r,\delta)$ condition is a sufficient condition to avoid such improperness.
However, the \emph{properness} of the layered locality parameters (and therefore the ordered $(r,\delta)$ condition)
is not a sufficient condition for the existence of relevant codes.

The notion of layered locality provides a useful property denoted in the following remark,
which makes the problem of analyzing the dimension and minimum distance characteristics of the codes more tractable.

\begin{remark} \label{rem:LL}
    A symbol with index $ i' \in \mathcal{N}_{j'} $ of codes with layered locality cannot participate in the repair process
    of another symbol with index $ i \in \mathcal{N}_j $ such that $ j < j' $,
    since we otherwise have a contradiction such that the $i'$th symbol has $(r_j,\delta_j)$-locality.
    In other words, the symbol index set $\mathcal{S}_i$ corresponding to the punctured code of the $i$th symbol satisfies
    $ \mathcal{S}_i \cap \bigsqcup_{j'=j+1}^{s^*} \mathcal{N}_{j'} = \emptyset $,
    hence $ \mathcal{S}_i \subset \bigsqcup_{j'=1}^{j} \mathcal{N}_{j'} $.
\end{remark}

Studying the characteristics of layered locality is not only important
as an intermediate step for solving the original problem of unequal locality,
but also can be justified by itself in some sense, under the ordered $(r,\delta)$ condition.
In particular, since smaller $r$ as well as larger $\delta$ are expected to allow a smaller distance $d$,
as suggested by \eqref{eq:PBND},
it is further expected that codes with large minimum distance are not ruled out by the layered locality restriction.
In other words, codes that violate the layered locality restriction such that
a symbol specified to have $(r_j,\delta_j)$-locality also has $(r_{j'},\delta_{j'})$-locality, $ j' \in [j-1] $,
where $ r_{j'} \leq r_j $ and $ \delta_{j'} \geq \delta_j $ by the ordered $(r,\delta)$ condition,
are expected to be of smaller distance.

\section{Upper Bounds Based on Layered Locality} \label{sect:LLBND}

In this section, both a dimension upper bound and a minimum distance upper bound for codes with layered locality are provided.
The upper bounds are derived using Algorithm \ref{alg:LL}, which is based on the original algorithm by \cite{Gopalan12TIT}
and subsequent modifications in the literature \cite{Kamath14TIT,Kadhe16ISIT,Zeh16ISIT}.

\begin{algorithm} [t]
    \caption{Used in the Proof of Lemma \ref{lem:alg:LL} and Lemma \ref{lem:LL-distbound}} \label{alg:LL}
    \begin{algorithmic}[1]
        \STATE Let $ \mathcal{Q}_0 = \bigsqcup_{j'=1}^{j-1}\mathcal{N}_{j'} $, $ l = 0 $
        \WHILE{$ \GRANK(\mathcal{Q}_l) < \GRANK( \bigsqcup_{j'=1}^j \mathcal{N}_{j'} ) $} \label{alg:LL:while}
            \STATE Pick any $ i \in \mathcal{N}_j \setminus \mathcal{Q}_l $
                such that $ \GRANK(\mathcal{Q}_l \sqcup \{i\}) > \GRANK(\mathcal{Q}_l) $ \label{alg:LL:pick}
            \STATE $ l= l + 1 $
            \STATE $ \mathcal{Q}_l = \mathcal{Q}_{l-1} \cup \mathcal{S}_i $
        \ENDWHILE
        \STATE $ L = l $
    \end{algorithmic}
\end{algorithm}

In the algorithm, $\mathcal{S}_i$ denotes the support of the punctured code
by which the $i$th symbol has $(r_j,\delta_j)$-locality.
The following lemma shows some key properties of the algorithm on codes with layered locality.

\begin{lemma} \label{lem:alg:LL}
    In Algorithm \ref{alg:LL}, we have
    \begin{enumerate}
        \item $ \lvert \mathcal{Q}_l \rvert - \lvert \mathcal{Q}_{l-1} \rvert
            \geq \GRANK(\mathcal{Q}_l) - \GRANK(\mathcal{Q}_{l-1}) + \delta_j - 1 $,
        \item $ L \geq \left\lceil \xi_j \mathbin{/} r_j \right\rceil $,
        \item $ \xi_j \leq k_j $,
    \end{enumerate}
    for $ l \in [L], j \in [s^*] $.
 \end{lemma}

\begin{proof}
    Due to Remark \ref{rem:LL}, we have $ \mathcal{S}_i \subset \bigsqcup_{j'=1}^j \mathcal{N}_{j'} $.
    Therefore, $ \mathcal{Q}_l \subset \bigsqcup_{j'=1}^j \mathcal{N}_{j'} $
    and $ \GRANK(\mathcal{Q}_l) \leq \GRANK( \bigsqcup_{j'=1}^j \mathcal{N}_{j'} ) $.
    The condition in Step \ref{alg:LL:while} ensures that
    it is always possible to pick a suitable index $i$ in Step \ref{alg:LL:pick}.
    The algorithm iterates until $ l = L $, where
    \begin{equation}
        \GRANK(\mathcal{Q}_L) = \GRANK( \bigsqcup_{j'=1}^j \mathcal{N}_{j'} ) \text{.} \label{eq:rank:LL}
    \end{equation}

    \begin{enumerate}
        \item First note that in the context of the punctured code with support $\mathcal{S}_i$,
            the symbols indexed by an arbitrary subset of $\mathcal{S}_i$ with the size of $ \delta_j - 1 $ are redundant
            since $ d(\mathscr{C}\rvert_{\mathcal{S}_i}) \geq \delta_j $.  
            We have $ \lvert \mathcal{Q}_l \rvert - \lvert \mathcal{Q}_{l-1} \rvert
                = \lvert \mathcal{Q}_l \setminus \mathcal{Q}_{l-1} \vert \geq \delta_j $,
            since otherwise we must have $ \GRANK(\mathcal{Q}_l) = \GRANK(\mathcal{Q}_{l-1}) $,
            due to the fact that $ \mathcal{Q}_l \setminus \mathcal{Q}_{l-1} \subset \mathcal{S}_i $,
            which is contradictory to the condition in Step \ref{alg:LL:pick}. 
            Now, out of the $ \lvert \mathcal{Q}_l \rvert - \lvert \mathcal{Q}_{l-1} \rvert \geq \delta_j $
            incremental symbols in the set $\mathcal{Q}_l$,
            at least $ \delta_j - 1 $ symbols are redundant since they are already redundant
            in the context of $ \mathcal{S}_i \subset \mathcal{Q}_l $.
            Therefore, we get
            \begin{equation*}
                \GRANK(\mathcal{Q}_l) - \GRANK(\mathcal{Q}_{l-1})
                    \leq \lvert \mathcal{Q}_l \rvert - \lvert \mathcal{Q}_{l-1} \rvert - ( \delta_j - 1 ) \text{.}
            \end{equation*}

        \item Since $ \mathcal{Q}_l = \mathcal{Q}_{l-1} \cup \mathcal{S}_i $, we have
            \begin{equation*}
                \GRANK(\mathcal{Q}_l) - \GRANK(\mathcal{Q}_{l-1}) \leq \GRANK(\mathcal{S}_i) \OVERSET{(a)}{\leq} r_j \text{,}
            \end{equation*}
            where (a) is due to Remark \ref{rem:LCSB}.
            This implies that
            \begin{equation*}
                L \geq \left\lceil \frac{ \GRANK(\mathcal{Q}_L) - \GRANK(\mathcal{Q}_0) }{r_j} \right\rceil \text{,}
            \end{equation*}
            and the second claim therefore directly follows from \eqref{eq:rank:LL}.

        \item Considering the incremental symbols in the construction of $ Q_L \subset \bigsqcup_{j'=1}^j \mathcal{N}_{j'} $
            we obtain
            \begin{align}
                n_j & \geq \lvert \mathcal{Q}_L \rvert - \lvert \mathcal{Q}_0 \rvert 
                    = \sum_{l=1}^{L} ( \lvert \mathcal{Q}_l \rvert - \lvert \mathcal{Q}_{l-1} \rvert ) \notag \\
                & \OVERSET{(a)}{\geq} \sum_{l=1}^{L} ( \GRANK(\mathcal{Q}_l) - \GRANK(\mathcal{Q}_{l-1}))
                    + L( \delta_j - 1 ) \notag \\
                & \OVERSET{(b)}{\geq}
                    \xi_j + \left\lceil \frac{\xi_j}{r_j} \right\rceil ( \delta_j - 1 ) \text{,}
                    \label{eq:lem:LL-rank}
            \end{align}
            where (a) comes from Lemma \ref{lem:alg:LL}-1), and (b) is due to Lemma \ref{lem:alg:LL}-2)
            and \eqref{eq:rank:LL}.
        
            For $ 0 \leq q_j \leq \delta_j - 2 $, suppose that $ \xi_j \geq p_j r_j + 1 $.
            It follows from \eqref{eq:lem:LL-rank} that
            \begin{align*}
                n_j & \geq p_j r_j + 1 + ( p_j + 1 )( \delta_j - 1 ) \\
                & = p_j ( r_j + \delta_j - 1 ) + \delta_j \\
                & > p_j ( r_j + \delta_j - 1 ) + q_j \\
                & = n_j \text{,}
            \end{align*}
            which is a contradiction. Therefore, we have
            \begin{equation*}
                \xi_j \leq p_j r_j = \lfloor m_j \rfloor r_j \text{.}
            \end{equation*}
        
            On the other hand, for $ \delta_j - 1 \leq q_j \leq r_j + \delta_j - 2 $,
            suppose that $ \xi_j \geq p_j r_j + q_j - ( \delta_j - 1 ) + 1 $,
            hence $ \xi_j \geq p_j r_j + 1 $.
            Again by \eqref{eq:lem:LL-rank}, we have
            \begin{align*}
                n_j & \geq p_j r_j + q_j - ( \delta_j - 1 ) + 1 + ( p_j + 1 )( \delta_j - 1 ) \\
                & = p_j ( r_j + \delta_j - 1 ) + q_j + 1 \\
                & > n_j \text{,}
            \end{align*}
            and therefore
            \begin{align*}
                \xi_j & \leq p_j r_j + q_j - ( \delta_j - 1 ) \\
                & = n_j - \lceil m_j \rceil ( \delta_j - 1 ) \text{.}
            \end{align*}

    \end{enumerate}
\end{proof}

The following proposition provides the dimension upper bound as a simple corollary to Lemma \ref{lem:alg:LL}-3).

\begin{proposition}[Dimension upper bound for codes with layered locality] \label{prop:LL-dimbound}
    The dimension of codes with layered locality is upper bounded by
    \begin{equation*}
        k \leq \sum_{j=1}^{s^*} k_j \text{.}
    \end{equation*}
\end{proposition}

\begin{proof}
    Clearly by Lemma \ref{lem:alg:LL}-3), we have $ k = \GRANK([n]) = \GRANK(\bigsqcup_{j=1}^{s^*} \mathcal{N}_j) =
        \sum_{j=1}^{s^*} \xi_j \leq \sum_{j=1}^{s^*} k_j $.
\end{proof}

The minimum distance upper bound is based on the lemma below, where the parameters $\xi_j$ appear in the expression.
They are subsequently eliminated in the proposition following the lemma.

\begin{lemma} \label{lem:LL-distbound}
    The minimum Hamming distance of codes with layered locality is upper bounded by
    \begin{equation*}
        d \leq n - k + 1 - \sum_{j=1}^{\sigma-1}(n_j - \xi_j) -
            \left(\left\lceil \frac{ k - \sum_{j=1}^{\sigma-1} \xi_j }{r_{\sigma}} \right\rceil - 1 \right)
            (\delta_\sigma - 1) \text{,}
    \end{equation*}
    where
    \begin{equation*}
        \sigma = \min\{ j \in [s^*] \mid \sum_{j'=1}^{j} \xi_{j'} = k \} \text{.}
    \end{equation*}
\end{lemma}

\begin{proof}
    Let us build a set $ \mathcal{T} \subset [n] $ such that $ \GRANK(\mathcal{T}) \leq k - 1 $,
    and apply Lemma \ref{lem:dist-red} to obtain the required distance upper bound.
    First, set $ j = \sigma $ in Algorithm \ref{alg:LL}.
    By Lemma \ref{lem:alg:LL}-2) and the definition of $\sigma$, we have
    \begin{equation*}
        L \geq \left\lceil \frac{\xi_{\sigma}}{r_{\sigma}} \right\rceil \geq
        \left\lceil \frac{ k - \sum_{j=1}^{\sigma-1} \xi_j }{r_{\sigma}} \right\rceil \text{.}
    \end{equation*}
    Let $ \mathcal{T} = \mathcal{Q}_l $ where
    \begin{equation*}
        l = \left\lceil \frac{ k - \sum_{j=1}^{\sigma-1} \xi_j }{r_{\sigma}} \right\rceil - 1 \text{.}
    \end{equation*}
    Since $ l \leq L - 1 $, we have
    \begin{align*}
        \GRANK(\mathcal{T}) & \leq \GRANK( \bigsqcup_{j=1}^\sigma \mathcal{N}_j ) - 1 = \sum_{j=1}^{\sigma} \xi_j - 1 \\
        & = k - 1 \text{.}
    \end{align*}
    We conclude the proof by noting that the number of redundant symbols indexed by $\mathcal{T}$ is
    \begin{align*}
        \gamma & = \lvert \mathcal{T} \rvert - \GRANK(\mathcal{T}) = \lvert \mathcal{Q}_l \rvert - \GRANK(\mathcal{Q}_l) \\
        & = \sum_{l'=1}^{l} ( \lvert \mathcal{Q}_{l'} \rvert - \lvert \mathcal{Q}_{l'-1} \rvert )
            + \lvert \mathcal{Q}_0 \rvert - \sum_{l'=1}^{l} ( \GRANK(\mathcal{Q}_{l'}) - \GRANK(\mathcal{Q}_{l'-1}) )
            - \GRANK(\mathcal{Q}_0) \\
        & \OVERSET{(a)}{\geq} \lvert \bigsqcup_{j=1}^{\sigma-1} \mathcal{N}_j \rvert
            - \GRANK( \bigsqcup_{j=1}^{\sigma-1} \mathcal{N}_j ) + l ( \delta_{\sigma} - 1 ) \\
        & = \sum_{j=1}^{\sigma-1} ( n_j - \xi_j ) + \left( \left\lceil \frac{ k - \sum_{j=1}^{\sigma-1} \xi_j }{r_{\sigma}}
            \right\rceil - 1 \right)( \delta_{\sigma} - 1 ) \text{,}
    \end{align*}
    where (a) is due to Lemma \ref{lem:alg:LL}-1).
\end{proof}

\begin{proposition}[Minimum distance upper bound for codes with layered locality] \label{prop:LL-distbound}
    The minimum Hamming distance of codes with layered locality is upper bounded by
    \begin{equation*}
        d \leq n - k + 1 - \sum_{j=1}^{s-1} ( n_j - k_j )
            - \left( \left\lceil \frac{ k - \sum_{j=1}^{s-1} k_j }{r_s} \right\rceil - 1 \right)
            ( \delta_s - 1 ) \text{,}
    \end{equation*}
    where
    \begin{equation*}
        s = \min{\{ j \in [s^*] \mid \sum_{j'=1}^j k_{j'} \geq k \}} \text{.}
    \end{equation*}
\end{proposition}

\begin{proof}
    First note that $s$ is well defined due to Proposition \ref{prop:LL-dimbound},
    and we have $ s \leq \sigma $ since $ \sum_{j=1}^\sigma k_j \geq \sum_{j=1}^\sigma \xi_j = k $.
    If $ s = \sigma $, it is easy to verify that the proposition holds
    by applying Lemma \ref{lem:alg:LL}-3) on Lemma \ref{lem:LL-distbound}.
    
    Otherwise, if $ s \leq \sigma - 1 $, we get
    \begin{align}
        d \OVERSET{(a)}{\leq} {}& n - k + 1 - \sum_{j=1}^{\sigma-1} ( n_j - \xi_j ) -
            \left( \left\lceil \frac{ k - \sum_{j=1}^{\sigma-1} \xi_j }{r_\sigma} \right\rceil - 1 \right)
            ( \delta_\sigma - 1 ) \notag \\
        \OVERSET{(b)}{\leq} {}& n - k + 1 - \sum_{j=1}^s ( n_j - \xi_j ) \notag \\
        \OVERSET{(c)}{\leq} {}& n - k + 1 - \sum_{j=1}^{s-1} ( n_j - k_j ) - ( n_s - k_s ) \text{,}
            \label{eq:prop:LL-distbound}
    \end{align}
    where (a) is just Lemma \ref{lem:LL-distbound}, (b) is obtained by removing some non-negative subtrahends,
    and (c) is due to Lemma \ref{lem:alg:LL}-3).
    Note that, if $ 0 \leq q_s \leq \delta_s - 2 $, we can write
    \begin{align}
        n_s - k_s & = n_s - \lfloor m_s \rfloor r_s \geq n_s - m_s r_s \notag \\
        & = m_s ( \delta_s - 1 ) \geq \lfloor m_s \rfloor( \delta_s - 1 ) \notag \\
        & = \frac{k_s}{r_s} ( \delta_s - 1 ) \text{.} \label{eq:prop:LL-distbound:1}
    \end{align}
    Otherwise, if $ \delta_s - 1 \leq q_s \leq r_s + \delta_s - 2 $, again we get
    \begin{align}
        n_s - k_s & = \lceil m_s \rceil ( \delta_s - 1 ) \geq \frac{m_s r_s}{r_s}( \delta_s - 1 ) \notag \\
        & = \frac{ n_s - m_s ( \delta_s - 1 ) }{r_s} ( \delta_s - 1 )
            \geq \frac{ n_s - \lceil m_s \rceil ( \delta_s - 1 ) }{r_s} ( \delta_s - 1 ) \notag \\
        & = \frac{k_s}{r_s} ( \delta_s - 1 ) \text{.} \label{eq:prop:LL-distbound:2}
    \end{align}
    Furthermore, we have
    \begin{equation}
        \frac{k_s}{r_s} \geq \frac{ k - \sum_{j=1}^{s-1} k_j }{r_s} >
            \left\lceil \frac{ k - \sum_{j=1}^{s-1} k_j }{r_s} \right\rceil - 1 \text{.} \label{eq:prop:LL-distbound:3}
    \end{equation}
    Therefore, substituting \eqref{eq:prop:LL-distbound:1}, \eqref{eq:prop:LL-distbound:2} and \eqref{eq:prop:LL-distbound:3}
    into \eqref{eq:prop:LL-distbound} completes the proof.
\end{proof}

For the conventional $r$-locality case, note that $ k_j = n_j - \lceil m_j \rceil $ regardless of $q_j$.
Further substituting $ r_j = j $ results in \eqref{eq:Kadhe}.

\section{Upper Bounds Based on Unequal Locality} \label{sect:ULBND}

The dimension and minimum distance upper bounds for codes with unequal locality are derived by
characterizing the relation between layered locality and unequal locality,
and applying the result on the bounds based on layered locality.

It is easy to see that codes having unequal locality with parameters $\{(n_j,r_j,\delta_j)\}_{j\in[s^*]}$
also have layered locality with parameters $\{(\hat{n}_j^*,r_j,\delta_j)\}_{j\in[s^*]}$
for some $\{\hat{n}_j^*\}_{j\in[s^*]}$.%
\footnote{
    We use a \emph{hat} notation hereafter, to denote layered locality parameters, such as
    $\hat{n}_j$, $\hat{p}_j$, $\hat{q}_j$, $\hat{m}_j$, $\hat{k}_j$, and $\hat{s}$,
    to distinguish them from ordinary unequal locality parameters.
}
Specifically, let $\mathcal{N}_j$, $ j \in [s^*] $, be the corresponding symbol index sets
for the unequal locality parameters of $\{(n_j,r_j,\delta_j)\}_{j\in[s^*]}$.
Define $\tilde{\mathcal{N}}_j$ to be the index set of all symbols having $(r_j,\delta_j)$-locality, and let
\begin{equation*}
    \hat{\mathcal{N}}_j^* = \tilde{\mathcal{N}}_j \setminus \bigcup_{j'=1}^{j-1} \tilde{\mathcal{N}}_{j'} \text{.}
\end{equation*}
It follows that every symbol with index $ i \in \hat{\mathcal{N}}_j^* $, $ j \in [s^*] $, has $(r_j,\delta_j)$-locality
but not $(r_{j'},\delta_{j'})$-locality such that $ j' \in [j-1] $.
Furthermore, for any $ j_1,j_2 \in [s^*] $, $ j_1 < j_2 $, we have
\begin{align*}
    \hat{\mathcal{N}}_{j_1}^* \cap \hat{\mathcal{N}}_{j_2}^* & \subset \tilde{\mathcal{N}}_{j_1} \cap
        ( \tilde{\mathcal{N}}_{j_2} \setminus \bigcup_{j'=1}^{j_2-1} \tilde{\mathcal{N}}_{j'} )
        = \tilde{\mathcal{N}}_{j_1} \cap
        ( \tilde{\mathcal{N}}_{j_2} \cap \bigcap_{j'=1}^{j_2-1} \tilde{\mathcal{N}}_{j'}^{\mathsf{c}} ) \\
    & = \tilde{\mathcal{N}}_{j_1} \cap \tilde{\mathcal{N}}_{j_1}^{\mathsf{c}} \cap \tilde{\mathcal{N}}_{j_2} \cap
        \bigcap_{j'\in[j_2-1]\setminus\{j_1\}} \tilde{\mathcal{N}}_{j'}^{\mathsf{c}} \\
    & = \emptyset \text{,}
\end{align*}
hence $ \hat{\mathcal{N}}_{j_1}^* \cap \hat{\mathcal{N}}_{j_2}^* = \emptyset $.
We also get
\begin{align}
    \bigsqcup_{j'=1}^j \hat{\mathcal{N}}_{j'}^* & = \tilde{\mathcal{N}}_1 \cup
        ( \tilde{\mathcal{N}}_2 \setminus \tilde{\mathcal{N}}_1 ) \cup
        ( \tilde{\mathcal{N}}_3 \setminus ( \tilde{\mathcal{N}}_1 \cup \tilde{\mathcal{N}}_2 ) ) \cup \cdots \notag \\
    & = ( \tilde{\mathcal{N}}_1 \cup \tilde{\mathcal{N}}_2 ) \cup
        ( \tilde{\mathcal{N}}_3 \setminus ( \tilde{\mathcal{N}}_1 \cup \tilde{\mathcal{N}}_2 ) ) \cup \cdots = \cdots
        = \bigcup_{j'=1}^j \tilde{\mathcal{N}}_{j'} \notag \\
    & \supset \bigsqcup_{j'=1}^j \mathcal{N}_{j'} \text{,} \label{eq:layered-n-set}
\end{align}
for all $ j \in [s^*] $, hence $ \bigsqcup_{j'=1}^{s^*} \hat{\mathcal{N}}_{j'}^* = [n] $.
This shows that the symbol index sets $\hat{\mathcal{N}}_j^*$, $ j \in [s^*] $,
define valid layered locality parameters of $\{(\hat{n}_j^*,r_j,\delta_j)\}_{j\in[s^*]}$,
where $ \hat{n}_j^* = \lvert \hat{\mathcal{N}}_j \rvert $.

Note that \eqref{eq:layered-n-set} implies that
\begin{equation}
    \sum_{j'=1}^j \hat{n}_{j'}^* \geq \sum_{j'=1}^j n_{j'} \text{,} \label{eq:layered-n}
\end{equation}
for all $ j \in [s^*] $,
which immediately yields the following upper bounds on codes with unequal locality
by maximizing the upper bounds based on layered locality over the relevant layered locality parameters.

\begin{remark} \label{rem:UL-bounds-LL}
    Recall that codes having unequal locality with parameters $\{(n_j,r_j,\delta_j)\}_{j\in[s^*]}$, also have layered locality.
    Denoting the layered locality parameters as $\{(\hat{n}^*_j,r_j,\delta_j)\}_{j\in[s^*]}$,
    we clearly have $ k \leq k_{UB}^{layered}(\hat{\mathbf{n}}^*) $ and $ d \leq d_{UB}^{layered}(\hat{\mathbf{n}}^*) $, 
    where $ \hat{\mathbf{n}}^* = (\hat{n}^*_1,\cdots,\hat{n}^*_{s^*}) $,
    and $k_{UB}^{layered}(\hat{\mathbf{n}}^*)$ and $d_{UB}^{layered}(\hat{\mathbf{n}}^*)$ are
    the dimension and the minimum distance upper bounds for codes having layered locality
    with parameters $\{(\hat{n}^*_j,r_j,\delta_j)\}_{j\in[s^*]}$
    (Proposition \ref{prop:LL-dimbound} and \ref{prop:LL-distbound})%
    \footnote{
        The remark is still valid for $k_{UB}^{layered}$ and $d_{UB}^{layered}$
        being any other dimension and minimum distance upper bounds for codes having layered locality.
    }
    , respectively.
    Let
    \begin{equation*}
        \mathcal{P} = \{ \hat{\mathbf{n}} \in \mathbb{Z}_{\scriptscriptstyle \geq 0}^{s^*} \mid
            \sum_{j'=1}^j \hat{n}_{j'} \geq \sum_{j'=1}^j n_{j'} \text{, } j \in [s^*] \} \text{.}
    \end{equation*}
    where $ \hat{\mathbf{n}} = (\hat{n}_1,\cdots,\hat{n}_{s^*}) $.
    Since $ \hat{\mathbf{n}}^* \in \mathcal{P} $, we clearly have%
    \footnote{
        It is possible that $d_{UB}^{layered}(\hat{\mathbf{n}})$ is undefined for some
        $ \hat{\mathbf{n}} \in \mathcal{P} $ (if, for example, $s$ in Proposition \ref{prop:LL-distbound} is undefined).
        It is assumed that such cases are discarded in the maximization.
    }
    \begin{align*}
        k & \leq \max_{ \hat{\mathbf{n}} \in \mathcal{P} } \left\{ k_{UB}^{layered}(\hat{\mathbf{n}}) \right\} \text{,} \\
        d & \leq \max_{ \hat{\mathbf{n}} \in \mathcal{P} } \left\{ d_{UB}^{layered}(\hat{\mathbf{n}}) \right\} \text{.}
    \end{align*}
\end{remark}

The upper bounds by the remark above are less desirable in that exhaustive maximization is required
over the set $\mathcal{P}$ of relevant layered localities which can be very large.
In the following, we derive upper bounds in closed form, given that the ordered $(r,\delta)$ condition holds.
First consider the lemma below.
Note that, in the proof, the main summand of the summation is always non-negative due to \eqref{eq:layered-n},
and the scaling in the denominator is therefore valid.
This technique will also be utilized several times in the proof the main theorem.

\begin{lemma} \label{lem:layered-n:cor}
    Consider codes having unequal locality with parameters $\{(n_j,r_j,\delta_j)\}_{j\in[s^*]}$ and ordered-$(r,\delta)$.
    For any layered locality with parameters $\{(\hat{n}_j,r_j,\delta_j)\}_{j\in[s^*]}$
    such that $ \hat{\mathbf{n}} = (\hat{n}_1,\cdots,\hat{n}_{s^*}) \in \mathcal{P} $,
    where $\mathcal{P}$ is given in Remark \ref{rem:UL-bounds-LL}, we have
    \begin{equation*}
        \sum_{j'=1}^j \hat{m}_{j'} ( \delta_{j'} - 1 ) \geq \sum_{j'=1}^j m_{j'} ( \delta_{j'} - 1 ) \text{,}
    \end{equation*}
    for all $ j \in [s^*] $.
\end{lemma}

\begin{proof}
    By repeatedly using \eqref{eq:layered-n} and the ordered $(r,\delta)$ condition, we have
    \begin{align*}
        \sum_{j'=1}^j ( \hat{m}_{j'} - m_{j'} )( \delta_{j'} - 1 )
            & = \frac{ \hat{n}_1 - n_1 }{ r_1 \mathbin{/} ( \delta_1 - 1 ) + 1 }
            + \frac{ \hat{n}_2 - n_2 }{ r_2 \mathbin{/} ( \delta_2 - 1 ) + 1 } 
            + \cdots + \frac{ \hat{n}_j - n_j }{ r_j \mathbin{/} ( \delta_j - 1 ) + 1 } \\
        & \geq \frac{ \sum_{j'=1}^2 ( \hat{n}_{j'} - n_{j'} ) }{ r_2 \mathbin{/} ( \delta_2 - 1 ) + 1 }
            + \cdots + \frac{ \hat{n}_j - n_j }{ r_j \mathbin{/} ( \delta_j - 1 ) + 1 } \\
        & \geq \cdots \\
        & \geq \frac{ \sum_{j'=1}^j ( \hat{n}_{j'} - n_{j'} ) }{ r_j \mathbin{/} ( \delta_j - 1 ) + 1 } \\
        & \geq 0 \text{.}
    \end{align*}
\end{proof}

The following two theorems, which are the main results of this section,
presents the dimension and minimum distance upper bounds
for codes having unequal locality with ordered-$(r,\delta)$ in closed form.

\begin{theorem}[Dimension upper bound for codes having unequal locality with ordered-$(r,\delta)$] \label{thm:UL-dimbound}
    The dimension of codes having unequal locality with ordered-$(r,\delta)$ is upper bounded by
    \begin{equation*}
        k \leq \sum_{j=1}^{s^*} m_j r_j \text{.}
    \end{equation*}
\end{theorem}

\begin{proof}
    Denoting the layered locality parameters of the codes as $\{(\hat{n}_j,r_j,\delta_j)\}_{j\in[s^*]}$, we can write
    \begin{align*}
        k & \OVERSET{(a)}{\leq} \sum_{j=1}^{s^*} \hat{k}_j \\
        & \OVERSET{(b)}{\leq} \sum_{j=1}^{s^*} \{ \hat{n}_j - \hat{m}_j ( \delta_j - 1 ) \}
            = n - \sum_{j=1}^{s^*} \hat{m}_j ( \delta_j - 1 ) \\
        & \OVERSET{(c)}{\leq} \sum_{j=1}^{s^*} n_j - \sum_{j=1}^{s^*} m_j ( \delta_j - 1 ) \\
        & = \sum_{j=1}^{s^*} m_j r_j \text{,}
    \end{align*}
    where (a) is Proposition \ref{prop:LL-dimbound}, (b) is from Definition \ref{def:LL},
    and (c) is due to Lemma \ref{lem:layered-n:cor}.
\end{proof}

Note that the dimension upper bound above characterizes the feasible rate region of
codes having unequal locality with ordered-$(r,\delta)$, and reduces to \eqref{eq:EL-dimbound} when $ s^* = 1 $.

\begin{theorem}[Minimum distance upper bound for codes having unequal locality with ordered-$(r,\delta)$]
\label{thm:UL-distbound}
    The minimum Hamming distance of codes having unequal locality with ordered-$(r,\delta)$ is upper bounded by
    \begin{equation*}
        d \leq n - k + 1 - \sum_{j=1}^{s-1} \lfloor m_j \rfloor ( \delta_j - 1 )
            - \left( \left\lceil \frac{ k - \sum_{j=1}^{s-1} \lfloor m_j \rfloor r_j }{r_s} \right\rceil - 1 \right)
            ( \delta_s - 1 ) \text{,}
    \end{equation*}
    where
    \begin{equation*}
        s = \max{\{ 0 \leq j \leq s^* - 1 \mid \sum_{j'=1}^j \lfloor m_{j'} \rfloor r_{j'} < k \}} + 1 \text{.}
    \end{equation*}
\end{theorem}

\begin{algorithm} [t]
    \caption{Used in the Proof of Theorem \ref{thm:UL-distbound}} \label{alg:jl}
    \begin{algorithmic}[1]
        \STATE Let $ j_0 = 0, l = 0, l' = 1 $
        \WHILE{$ l' \leq \hat{s} - 1 $}
            \IF{$ \sum_{j=j_l+1}^{l'} ( \hat{p}_j - p_j + \hat{\phi}_j ) < 0 $}
                \STATE $ l = l + 1 $
                \STATE $ j_l = l' $
            \ENDIF
            \STATE $ l' = l' + 1 $
        \ENDWHILE
        \STATE $ L = l $
    \end{algorithmic}
\end{algorithm}

\begin{proof}
    Recall that the codes also have layered locality with parameters $\{(\hat{n}_j,r_j,\delta_j)\}_{j\in[s^*]}$.
    By Proposition \ref{prop:LL-distbound}, we have
    \begin{equation}
        d \leq n - k + 1 - \sum_{j=1}^{\hat{s}-1} (\hat{n}_j - \hat{k}_j)
            - \left( \left\lceil \frac{ k - \sum_{j=1}^{\hat{s}-1} \hat{k}_j }{r_{\hat{s}}} \right\rceil - 1 \right)
            (\delta_{\hat{s}} - 1) \text{.} \label{eq:thm:UL-distbound:layered}
    \end{equation}
    For $ j \in [s^*] $, let
    \begin{equation*}
        \hat{\phi}_j = \begin{cases}
            0 & \text{if $ 0 \leq \hat{q}_j \leq \delta_j - 2 $,}\\
            1 & \text{if $ \delta_j - 1 \leq \hat{q}_j \leq r_j + \delta_j - 2 $.}
        \end{cases}
    \end{equation*}
    Note that
    \begin{equation}
        \hat{k}_j = \hat{p}_j r_j + ( \hat{q}_j - \delta_j + 1 )\hat{\phi}_j \text{,} \label{eq:k:1}
    \end{equation}
    and also
    \begin{equation}
        \hat{n}_j - \hat{k}_j = ( \hat{p}_j + \hat{\phi}_j )( \delta_j - 1 ) + \hat{q}_j ( 1 - \hat{\phi}_j )
            \text{.} \label{eq:k:2}
    \end{equation}
    The proof will proceed with the corresponding cases.

    \emph{Case 1}: $ s \geq \hat{s} $.\\
    Substituting \eqref{eq:k:1} and \eqref{eq:k:2} into \eqref{eq:thm:UL-distbound:layered} yields
    \begin{align}
        d \leq {}& n - k + 1 -
            \sum_{j=1}^{\hat{s}-1} \{ ( \hat{p}_j + \hat{\phi}_j )( \delta_j - 1 ) + \hat{q}_j ( 1 - \hat{\phi}_j ) \}
            - \left( \left\lceil X \right\rceil - 1 \right)( \delta_{\hat{s}} - 1 ) \notag \\
        \leq {}& n - k + 1 - \sum_{j=1}^{\hat{s}-1} p_j ( \delta_j - 1 ) - Y \text{,} \label{eq:thm:UL-distbound:1}
    \end{align}
    with
    \begin{align*}
        X & = \frac{ k - \sum_{j=1}^{\hat{s}-1} \{ \hat{p}_j r_j + ( \hat{q}_j - \delta_j + 1 ) \hat{\phi}_j \} }{r_{\hat{s}}}
            \text{,} \\
        Y & = \left( \left\lceil \frac{ k - \sum_{j=1}^{\hat{s}-1} p_j r_j }{r_{\hat{s}}} \right\rceil - 1
            + \lfloor A \rfloor + B \right)( \delta_{\hat{s}} - 1 ) \text{,}
    \end{align*}
    \begin{align}
        A & = \frac{ - \sum_{j=1}^{\hat{s}-1} \{ ( \hat{p}_j - p_j ) r_j +
        ( \hat{q}_j -\delta_j + 1 ) \hat{\phi}_j \} }{r_{\hat{s}}} +
            \sum_{j=1}^{\hat{s}-1} ( \hat{p}_j - p_j + \hat{\phi}_j ) \notag \\
        & = \frac{ \sum_{j=1}^{\hat{s}-1} ( \hat{p}_j - p_j )( r_{\hat{s}} - r_j ) +
            \sum_{j=1}^{\hat{s}-1} ( r_{\hat{s}} + \delta_j - 1 - \hat{q}_j )\hat{\phi}_j }{r_{\hat{s}}} \label{eq:A}
    \end{align}
    and
    \begin{align}
        B & = \frac{ \sum_{j=1}^{\hat{s}-1} \{ ( \hat{p}_j - p_j + \hat{\phi}_j )( \delta_j - 1 ) +
            \hat{q}_j ( 1 - \hat{\phi}_j ) \} }{ \delta_{\hat{s}} - 1 } -
            \sum_{j=1}^{\hat{s}-1} ( \hat{p}_j - p_j + \hat{\phi}_j ) \notag \\
        & = \frac{ \sum_{j=1}^{\hat{s}-1} ( \hat{p}_j - p_j + \hat{\phi}_j )( \delta_j - \delta_{\hat{s}} ) +
            \sum_{j=1}^{\hat{s}-1} \hat{q}_j ( 1 - \hat{\phi}_j ) }{ \delta_{\hat{s}} - 1 } \text{,} \label{eq:B}
    \end{align}
    where \eqref{eq:thm:UL-distbound:1} follows from the fact that
    $ \lceil a - b \rceil \geq \lceil a \rceil + \lfloor -b \rfloor $.

    Next, we will show that $ \lfloor A \rfloor + B \geq 0 $.
    Define $j_l$, $ 0 \leq l \leq L \leq \hat{s} - 1 $, according to Algorithm \ref{alg:jl}.
    The terms $ \hat{p}_j - p_j + \hat{\phi}_j $ starting from $ j = j_{l-1} + 1 $ are accumulated
    while the summation remains non-negative, and $ j_l $ is defined accordingly as the summation becomes negative.
    Note that, $ j_0 = 0 < j_1 < \cdots < j_L \leq \hat{s} - 1 $ such that, for $ l \in [L] $,
    \begin{equation}
        \sum_{j=j_{l-1}+1}^{j'} ( \hat{p}_j - p_j + \hat{\phi}_j ) \geq 0 \text{,} \label{eq:A:cond:1}
    \end{equation}
    $ j_{l-1} + 1 \leq j' \leq j_l - 1 $, and
    \begin{equation}
        \sum_{j=j_{l-1}+1}^{j_l} ( \hat{p}_j - p_j + \hat{\phi}_j ) < 0 \text{.} \label{eq:A:cond:2}
    \end{equation}
    Also, we have
    \begin{equation}
        \sum_{j=j_L+1}^{j'} ( \hat{p}_j - p_j + \hat{\phi}_j ) \geq 0 \text{,} \label{eq:A:cond:3}
    \end{equation}
    $ j_L + 1 \leq j' \leq \hat{s} - 1 $.
    Starting from \eqref{eq:A}, we can write
    \begin{align*}
        A & \OVERSET{(a)}{\geq}
            \frac{ \sum_{j=1}^{\hat{s}-1} ( \hat{p}_j - p_j + \hat{\phi}_j )( r_{\hat{s}} - r_j ) }{ r_{\hat{s}} } \\
        & = \frac{ \sum_{l=1}^L \sum_{j=j_{l-1}+1}^{j_l} ( \hat{p}_j - p_j + \hat{\phi}_j )( r_{\hat{s}} - r_j ) }
            {r_{\hat{s}}}
            + \frac{ \sum_{j=j_L+1}^{\hat{s}-1} ( \hat{p}_j - p_j + \hat{\phi}_j )( r_{\hat{s}} - r_j ) }{r_{\hat{s}}} \\
        & \OVERSET{(b)}{\geq}
            \frac{ \sum_{l=1}^L \sum_{j=j_{l-1}+1}^{j_l} ( \hat{p}_j - p_j + \hat{\phi}_j )( r_{\hat{s}} - r_{j_l} ) }
            {r_{\hat{s}}}
            + \frac{ \sum_{j=j_L+1}^{\hat{s}-1} ( \hat{p}_j - p_j + \hat{\phi}_j )( r_{\hat{s}} - r_{\hat{s}-1} ) }
            {r_{\hat{s}}} \\
        & \geq \frac{ \sum_{l=1}^L ( r_{\hat{s}} - r_{j_l} ) \sum_{j=j_{l-1}+1}^{j_l} ( \hat{p}_j - p_j + \hat{\phi}_j ) }
            {r_{\hat{s}}} \\
        & \OVERSET{\eqref{eq:A:cond:2}}{\geq} \sum_{l=1}^L \sum_{j=j_{l-1}+1}^{j_l} ( \hat{p}_j - p_j + \hat{\phi}_j ) \\
        & = \sum_{j=1}^{j_L} ( \hat{p}_j - p_j + \hat{\phi}_j ) \text{,}
    \end{align*}
    where (a) is due to the fact that $ \hat{q}_j < r_j + \delta_j - 1 $, and (b) follows from \eqref{eq:A:cond:1}
    and \eqref{eq:A:cond:3} with a derivation similar to Lemma \ref{lem:layered-n:cor}.
    As an intermediate result, we get
    \begin{equation*}
        \lfloor A \rfloor \geq \sum_{j=1}^{j_L} ( \hat{p}_j - p_j + \hat{\phi}_j ) \text{.}
    \end{equation*}

    As for B, first note that \eqref{eq:layered-n} implies that
    \begin{equation*}
        \sum_{j=1}^{j_L} \hat{q}_j \geq \sum_{j=1}^{j_L} q_j - \sum_{j=1}^{j_L} ( \hat{p}_j - p_j )( r_j + \delta_j - 1 )
            \text{,}
    \end{equation*}
    and therefore
    \begin{align}
        \sum_{j=1}^{j_L} \hat{q}_j ( 1 - \hat{\phi}_j ) & \geq
            \sum_{j=1}^{j_L} q_j - \sum_{j=1}^{j_L} ( \hat{p}_j - p_j )( r_j + \delta_j - 1 ) -
            \sum_{j=1}^{j_L} \hat{q}_j \hat{\phi}_j \notag \\
        & \OVERSET{(a)}{\geq} - \sum_{j=1}^{j_L} ( \hat{p}_j - p_j + \hat{\phi}_j )( r_j + \delta_j - 1 ) \text{,}
            \label{eq:B:part}
    \end{align}
    where (a) holds since $ \hat{q}_j < r_j + \delta_j - 1 $.
    From \eqref{eq:B}, we can write
    \begin{align*}
        B & \OVERSET{(a)}{\geq} \frac{ \sum_{j=1}^{j_L} ( \hat{p}_j - p_j + \hat{\phi}_j )( \delta_j - \delta_{\hat{s}} )
            + \sum_{j=j_L+1}^{\hat{s}-1} ( \hat{p}_j - p_j + \hat{\phi}_j )( \delta_j - \delta_{\hat{s}} )}
            { \delta_{\hat{s}} - 1 } + \frac{\sum_{j=1}^{j_L} \hat{q}_j ( 1 - \hat{\phi}_j ) }{ \delta_{\hat{s}} - 1 } \\
        & \OVERSET{\eqref{eq:B:part}}{\geq} - \frac{ \sum_{l=1}^L \sum_{j=j_{l-1}+1}^{j_l}
            ( \hat{p}_j - p_j + \hat{\phi}_j )( r_j + \delta_{\hat{s}} - 1 )}{ \delta_{\hat{s}} - 1 }
            + \frac{ \sum_{j=j_L+1}^{\hat{s}-1} ( \hat{p}_j - p_j + \hat{\phi}_j )( \delta_j - \delta_{\hat{s}} ) }
            { \delta_{\hat{s}} - 1 } \\
        & \OVERSET{(b)}{\geq} - \frac{ \sum_{l=1}^L ( r_{j_l} + \delta_{\hat{s}} - 1 ) \sum_{j=j_{l-1}+1}^{j_l}
            ( \hat{p}_j - p_j + \hat{\phi}_j ) }{ \delta_{\hat{s}} - 1 } \\
        & \OVERSET{\eqref{eq:A:cond:2}}{\geq} - \sum_{j=1}^{j_L} ( \hat{p}_j - p_j + \hat{\phi}_j ) \text{,}
    \end{align*}
    where (a) is obtained by removing some non-negative subtrahends,
    and (b) comes from \eqref{eq:A:cond:1} and \eqref{eq:A:cond:3} with a derivation similar to Lemma \ref{lem:layered-n:cor}.

    Since $ \lfloor A \rfloor + B \geq 0 $, we have, continuing from \eqref{eq:thm:UL-distbound:1},
    \begin{align*}
        d & \leq n - k + 1 - \sum_{j=1}^{\hat{s}-1} p_j ( \delta_j - 1 ) - \left( \left\lceil
            \frac{ k - \sum_{j=1}^{\hat{s}-1} p_j r_j }{r_{\hat{s}}} \right\rceil - 1 \right)( \delta_{\hat{s}} - 1 ) \\
        & \leq n - k + 1 - \sum_{j=1}^{s-1} p_j ( \delta_j - 1 ) - \left( \left\lceil
            \frac{ k - \sum_{j=1}^{\hat{s}-1} p_j r_j - \sum_{j=\hat{s}}^{s-1} p_j r_{\hat{s}} }{r_{\hat{s}}}
            \right\rceil - 1 \right)( \delta_{\hat{s}} - 1 ) \\
        & \leq n - k + 1 - \sum_{j=1}^{s-1} p_j ( \delta_j - 1 ) - \left( \left\lceil
            \frac{ k - \sum_{j=1}^{s-1} p_j r_j }{r_s} \right\rceil - 1 \right)( \delta_s - 1 ) \\
        & = n - k + 1 - \sum_{j=1}^{s-1} \lfloor m_j \rfloor ( \delta_j - 1 ) - \left( \left\lceil
            \frac{ k - \sum_{j=1}^{s-1} \lfloor m_j \rfloor r_j }{r_s} \right\rceil - 1 \right)( \delta_s - 1 ) \text{.}
    \end{align*}

    \emph{Case 2}: $ s < \hat{s} $.\\
    It is easy to verify that \eqref{eq:k:2} implies $ \hat{n}_j - \hat{k}_j \geq \hat{m}_j ( \delta_j - 1 )$, $ j \in [s^*] $.
    By applying Lemma \ref{lem:layered-n:cor}, we can write
    \begin{equation}
        \sum_{j=1}^s ( \hat{n}_j - \hat{k}_j ) \geq \sum_{j=1}^s \hat{m}_j ( \delta_j - 1 ) \geq
            \sum_{j=1}^s  \lfloor m_j \rfloor ( \delta_j - 1 ) \text{.} \label{eq:nmk}
    \end{equation}
    On the other hand, since $ s \leq \hat{s} - 1 \leq s^* - 1 $,
    we have $ \sum_{j=1}^s \lfloor m_j \rfloor r_j \geq k $, leading to
    \begin{equation}
        \lfloor m_s \rfloor \geq \frac{ k - \sum_{j=1}^{s-1} \lfloor m_j \rfloor r_j }{r_s}
            > \left\lceil \frac{ k - \sum_{j=1}^{s-1} \lfloor m_j \rfloor r_j }{r_s} \right\rceil - 1 \text{.} \label{eq:mr}
    \end{equation}
    Therefore, from \eqref{eq:thm:UL-distbound:layered}, we get
    \begin{align*}
        d & \OVERSET{(a)}{\leq} n - k + 1 - \sum_{j=1}^s ( \hat{n}_j - \hat{k}_j ) \\
        & \OVERSET{(b)}{<} n - k + 1 - \sum_{j=1}^{s-1} \lfloor m_j \rfloor ( \delta_j - 1 ) -
            \left( \left\lceil \frac{ k - \sum_{j=1}^{s-1} \lfloor m_j \rfloor r_j }{r_s} \right\rceil - 1 \right)
            ( \delta_s - 1 ) \text{,}
    \end{align*}
    where (a) is obtained by removing some non-negative subtrahends, and (b) follows from \eqref{eq:nmk} and \eqref{eq:mr}.
\end{proof}

It is easy to verify that Theorem \ref{thm:UL-distbound} degenerates to \eqref{eq:PBND} when $ s^* = 1 $.

The next two corollaries show the relationship between the upper bounds
based on the exhaustive maximization (Remark \ref{rem:UL-bounds-LL})
and the upper bounds in closed form (Theorem \ref{thm:UL-dimbound} and \ref{thm:UL-distbound}),
given that the ordered $(r,\delta)$ condition is satisfied .
In particular, Corollary \ref{cor:UL-bndcmp} shows that the upper bounds in closed form are looser compared to
the upper bounds based on exhaustive maximization.
However, Corollary \ref{cor:RBC} characterizes an unequal locality parameter regime
where both types of the bounds coincide.

\begin{corollary} \label{cor:UL-bndcmp}
    For the dimension and minimum distance bounds of codes having unequal locality with ordered-$(r,\delta)$
    (Remark \ref{rem:UL-bounds-LL}, Theorem \ref{thm:UL-dimbound}, and Theorem \ref{thm:UL-distbound}), we have
    \begin{align*}
        k_{UB}^{Rem\text{-}\ref{rem:UL-bounds-LL}} & \leq k_{UB}^{Thm\text{-}\ref{thm:UL-dimbound}} \text{,} \\
        d_{UB}^{Rem\text{-}\ref{rem:UL-bounds-LL}} & \leq d_{UB}^{Thm\text{-}\ref{thm:UL-distbound}} \text{.}
    \end{align*}
\end{corollary}

\begin{proof}
    Let the unequal and layered locality parameters of the codes be $\{(n_j,r_j,\delta_j)\}_{j\in[s^*]}$ and
    $\{(\hat{n}_j^*,r_j,\delta_j)\}_{j\in[s^*]}$, respectively.
    Denoting $ \hat{\mathbf{n}}^* = (\hat{n}_1^*,\cdots,\hat{n}_{s^*}^*) $,
    the proofs of Theorem \ref{thm:UL-dimbound} and \ref{thm:UL-distbound} show that
    \begin{align*}
        k_{UB}^{layered} (\hat{\mathbf{n}}^*) \leq k_{UB}^{Thm\text{-}\ref{thm:UL-dimbound}} (\mathbf{n}) \text{,} \\
        d_{UB}^{layered} (\hat{\mathbf{n}}^*) \leq d_{UB}^{Thm\text{-}\ref{thm:UL-distbound}} (\mathbf{n}) \text{,}
    \end{align*}
    where $k_{UB}^{layered}(\hat{\mathbf{n}}^*)$ and $d_{UB}^{layered}(\hat{\mathbf{n}}^*)$
    are the dimension and the minimum distance upper bounds given by
    Proposition \ref{prop:LL-dimbound} and \ref{prop:LL-distbound}, respectively.
    Furthermore, identical derivations can be made for arbitrary $ \hat{\mathbf{n}} = (\hat{n}_1,\cdots,\hat{n}_{s^*}) $
    such that $ \hat{\mathbf{n}} \in \mathcal{P} $,
    where $\mathcal{P}$ is defined in Remark \ref{rem:UL-bounds-LL}.
    Therefore, we can write
    \begin{align*}
        k_{UB}^{layered} (\hat{\mathbf{n}}) \leq k_{UB}^{Thm\text{-}\ref{thm:UL-dimbound}} (\mathbf{n}) \text{,} \\
        d_{UB}^{layered} (\hat{\mathbf{n}}) \leq d_{UB}^{Thm\text{-}\ref{thm:UL-distbound}} (\mathbf{n}) \text{,}
    \end{align*}
    where $k_{UB}^{layered}(\hat{\mathbf{n}})$ and $d_{UB}^{layered}(\hat{\mathbf{n}})$ are the bounds by
    Proposition \ref{prop:LL-dimbound} and \ref{prop:LL-distbound}, respectively,
    on codes having layered locality with parameters $\{(\hat{n}_j,r_j,\delta_j)\}_{j\in[s^*]}$.
    Recalling that Remark \ref{rem:UL-bounds-LL} involves a maximization over all $ \hat{\mathbf{n}} \in \mathcal{P} $
    in the left hand sides, the proof is complete.
\end{proof}

\begin{corollary} \label{cor:RBC}
    For codes having unequal locality with ordered-$(r,\delta)$ such that $ r_j + \delta_j - 1 \mid n_j $, $ j \in [s^*] $,
    we have
    \begin{align*}
        k_{UB}^{Rem\text{-}\ref{rem:UL-bounds-LL}} & = k_{UB}^{Thm\text{-}\ref{thm:UL-dimbound}} \text{,} \\
        d_{UB}^{Rem\text{-}\ref{rem:UL-bounds-LL}} & = d_{UB}^{Thm\text{-}\ref{thm:UL-distbound}} \text{.}
    \end{align*}
\end{corollary}

\begin{proof}
    Let $k_{UB}^{layered}(\hat{\mathbf{n}})$ and $d_{UB}^{layered}(\hat{\mathbf{n}})$,
    where $ \hat{\mathbf{n}} = (\hat{n}_1,\cdots,\hat{n}_{s^*}) $,
    be the dimension and the minimum distance upper bounds for codes having layered locality
    with parameters $\{(\hat{n}_j,r_j,\delta_j)\}_{j\in[s^*]}$,
    given by Proposition \ref{prop:LL-dimbound} and \ref{prop:LL-distbound}, respectively.
    Furthermore, let $ \hat{\mathbf{n}} = \mathbf{n} = (n_1,\cdots,n_{s^*}) $.
    Then, we have
    \begin{align*}
        k_{UB}^{layered} (\hat{\mathbf{n}}) \OVERSET{(a)}{\leq} k_{UB}^{Rem\text{-}\ref{rem:UL-bounds-LL}} (\mathbf{n})
            \OVERSET{(b)}{\leq} k_{UB}^{Thm\text{-}\ref{thm:UL-dimbound}} (\mathbf{n}) \text{,} \\
        d_{UB}^{layered} (\hat{\mathbf{n}}) \OVERSET{(a)}{\leq} d_{UB}^{Rem\text{-}\ref{rem:UL-bounds-LL}} (\mathbf{n})
            \OVERSET{(b)}{\leq} d_{UB}^{Thm\text{-}\ref{thm:UL-distbound}} (\mathbf{n}) \text{,}
    \end{align*}
    where (a) is due to the definitions of $k_{UB}^{Rem\text{-}\ref{rem:UL-bounds-LL}}$ and
    $d_{UB}^{Rem\text{-}\ref{rem:UL-bounds-LL}}$, and (b) is just Corollary \ref{cor:UL-bndcmp}.
    The proof is complete by verifying that
    \begin{align*}
        k_{UB}^{layered} (\hat{\mathbf{n}}) & = k_{UB}^{layered} (\mathbf{n})
            = k_{UB}^{Thm\text{-}\ref{thm:UL-dimbound}} (\mathbf{n}) \text{,} \\
        d_{UB}^{layered} (\hat{\mathbf{n}}) & = d_{UB}^{layered} (\mathbf{n})
            = d_{UB}^{Thm\text{-}\ref{thm:UL-distbound}} (\mathbf{n}) \text{,}
    \end{align*}
    under the condition of $ r_j + \delta_j - 1 \mid n_j $, $ j \in [s^*] $,
    where $ r_j + \delta_j - 1 \mid n_j $ implies that $ m_j = \lceil m_j \rceil = \lfloor m_j \rfloor $
    and $ k_j = n_j - m_j ( \delta - 1 ) = m_j j $.
    Note that, in showing the equality between $d_{UB}^{layered}(\mathbf{n})$ and
    $d_{UB}^{Thm\text{-}\ref{thm:UL-distbound}}(\mathbf{n})$, we have
    \begin{align*}
        s^{Thm\text{-}\ref{thm:UL-distbound}}(\mathbf{n}) & =
            \max{\{ 0 \leq j \leq s^* - 1 \mid \sum_{j'=1}^j m_{j'} r_{j'} < k \}} + 1 \\
        & \OVERSET{(a)}{=} \min{\{ j \in [s^*] \mid \sum_{j'=1}^j m_{j'} r_{j'} \geq k \}} \\
        & = s^{layered}(\mathbf{n}) \text{,}
    \end{align*}
    where (a) is due to Theorem \ref{thm:UL-dimbound}.
\end{proof}

\section{Optimal Code Construction} \label{sect:construction}

We give an optimal code construction achieving the equality for the bound in Theorem \ref{thm:UL-distbound},
and also the distance bound in Remark \ref{rem:UL-bounds-LL} under the ordered $(r,\delta)$ condition.
In other words, the code is built in the parameter regime where the two distance bounds coincide,
as shown in Corollary \ref{cor:RBC}.
The construction closely follows the Gabidulin-based LRC construction
which originates from \cite{Silberstein13ISIT}, and is also used in \cite{Kadhe16ISIT}.

\begin{construction}[Gabidulin-based LRC with unequal locality] \label{cnstrct}
    For integers $ m_j \geq 1 $, $ r_j \geq 1 $, and $ \delta_j \geq 2 $, $ j \in [s^*] $,
    let $ n_j = m_j ( r_j + \delta_j - 1 ) $ and $ n = \sum_{j=1}^{s^*} n_j $.
    Let us also constrain the parameters to satisfy the condition $ k \leq \sum_{j=1}^{s^*} m_j r_j \leq t $.
    Linear $[n,k]_{q^t}$ codes are constructed according to the following steps.
    \begin{enumerate}
        \item Precode $k$ information symbols using a $[\sum_{j=1}^{s^*} m_j r_j,k]_{q^t}$ Gabidulin code.
        \item Partition the Gabidulin codeword symbols into $ \sum_{j=1}^{s^*} m_j $ local groups,
            where each of the $m_j$ groups is of size $r_j$, $ j \in [s^*] $.
        \item Encode each local group of size $r_j$ using a linear $[r_j+\delta_j-1,r_j,\delta_j]_q$ MDS code.%
            \footnote{
                The encoding is performed by multiplying the symbol vector corresponding to each local group
                of a Gabidulin codeword ($\mathbb{F}_{q^t}^{r_j}$) by the MDS generator matrix
                ($\mathbb{F}_q^{r_j\times(r_j+\delta_j-1)}$),
                where the actual scalar multiplication is over $\mathbb{F}_{q^t}$.
            }
    \end{enumerate}
\end{construction}

It is obvious by construction that a Gabidulin-based $(r,\delta)$-LRC $\mathscr{C}$
has indeed unequal locality with parameters $\{(n_j,r_j,\delta_j)\}_{j\in[s^*]}$.
In particular, by choosing $\mathcal{S}_i$ as the support of the MDS local code corresponding to the $i$th symbol,
we have $ i \in S_i $ and $ \mathcal{S}_i = r_j + \delta_j - 1 $.
Furthermore, $ d(\mathscr{C}\rvert_{\mathcal{S}_i}) \geq \delta_j $ since
$\mathscr{C}\rvert_{\mathcal{S}_i}$ is a subcode of an $[r_j+\delta_j-1,r_j,\delta_j]_q$ MDS code. 

Note that, by having $ k = \sum_{j=1}^{s^*} m_j r_j $ in the construction,
the equality in the dimension bound by Theorem \ref{thm:UL-dimbound} is achieved, showing its tightness.

We require the following remark and lemma to analyze the minimum distance of the code by Construction \ref{cnstrct}
with ordered-$(r,\delta)$, which is shown to be optimal in the theorem following the lemma.

\begin{remark} \label{rem:dsum}
    Clearly, by Lemma \ref{lem:ERANK},
    the subspace generated by the evaluation points of the code of Construction \ref{cnstrct}
    is a direct sum of each subspace generated by the evaluation points corresponding to a single local group.
    Therefore, $\ERANK(\mathcal{T})$ of some set $ \mathcal{T} \subset [n] $ is the sum of each $\ERANK(\cdot)$
    computed separately on the points in the same local group.
\end{remark}

\begin{lemma}\label{lem:rnkera}
    Let the parameters $\{(n_j,r_j,\delta_j)\}_{j\in[s^*]}$ in Construction \ref{cnstrct}
    satisfy the ordered $(r,\delta)$ condition.
    Suppose an ordered set $ \mathcal{L} = \{ \mathcal{G}_1, \ldots, \mathcal{G}_{\lvert \mathcal{L} \rvert } \} $
    such that $ \lvert \mathcal{L} \rvert = \sum_{j=1}^{s^*} m_j $,
    where each element of $\mathcal{L}$ is a symbol index set
    corresponding to the symbols of a distinct encoded local group in Construction \ref{cnstrct},
    and the order is according to the ordered $(r,\delta)$ condition.
    Elements of identical $(r,\delta)$ are ordered arbitrarily.
    Let denote an erasure pattern of $e$ erased symbols by the index set of the $ n - e $ remaining symbols.
    The index set $ \mathcal{R}^* \subset [n] $ of the $ n - e $ remaining symbols
    where the indices are taken greedily starting from the first element $\mathcal{G}_1$ of $\mathcal{L}$,
    corresponds to a worst case erasure pattern in terms of rank erasures (or remaining rank), i.e., we have
    \begin{equation*}
        \ERANK(\mathcal{R}) \geq \ERANK(\mathcal{R}^*) \text{,}
    \end{equation*}
    for any symbol index set $ \mathcal{R} \subset [n] $ such that $ \lvert \mathcal{R} \rvert = n - e $.
\end{lemma}

\begin{IEEEproof}
    See Appendix \ref{sect:appx:rnkera}.
\end{IEEEproof}

\begin{theorem}[Optimality of Gabidulin-based LRC with unequal locality and ordered-$(r,\delta)$] \label{thm:optOL}
    Gabidulin-based LRCs with unequal locality satisfying the ordered $(r,\delta)$ condition are distance optimal
    with respect to the distance upper bound for codes having unequal locality with ordered-$(r,\delta)$
    (Theorem \ref{thm:UL-distbound}).
\end{theorem}

\begin{proof}
    We derive a lower bound on the minimum distance of the code,
    which equals the upper bound of Theorem \ref{thm:UL-distbound}.
    In particular, we show that erasure correction is possible from an arbitrary symbol set with the cardinality of
    \begin{equation*}
        \tau = k + \sum_{j=1}^{s-1} m_j ( \delta_j - 1 ) +
            \left( \left\lceil \frac{ k - \sum_{j=1}^{s-1} m_j r_j }{r_s} \right\rceil - 1 \right)
            ( \delta_s - 1 ) \text{,}
    \end{equation*}
    where $s$ is given by Theorem \ref{thm:UL-distbound}.
    Applying Lemma \ref{lem:dist-rank:cor} with Remark \ref{rem:rank} gives the desired lower bound.
    
    Let integers $P$ and $Q$ such that
    \begin{equation}
        k - 1 - \sum_{j=1}^{s-1} m_j r_j = P r_s + Q \geq 0 \label{eq:thm:optOL:pqdef}
    \end{equation}
    and $ 0 \leq Q \leq r_s - 1 $.
    Consider an arbitrary symbol index set $ \mathcal{T} \subset [n] $ of cardinality
    \begin{equation}
        \lvert \mathcal{T} \rvert = \sum_{j=1}^{s-1} n_j + P ( r_s + \delta_s - 1 ) + Q + 1 \text{.}
            \label{eq:thm:optOL:Tcard}
    \end{equation}
    Let $\mathcal{T}^*$ be the greedily chosen set of Lemma \ref{lem:rnkera}
    such that $ \lvert \mathcal{T}^* \rvert = \lvert \mathcal{T} \rvert $.
    Then, $\mathcal{T}^*$ consists of all the symbols in the local groups of $(r_j,\delta_j)$, $j \in [s-1] $,
    $P$ local groups of $(r_s,\delta_s)$,
    and some $ Q + 1 $ symbols in an additional local group of $(r_s,\delta_s)$.
    This composition is valid since
    \begin{align*}
        P & \OVERSET{\eqref{eq:thm:optOL:pqdef}}{=} \frac{ k - \sum_{j=1}^{s-1} m_j r_j }{r_s}
            - \frac{ 1 + Q }{r_s} \\
        & \OVERSET{(a)}{\leq} \frac{ m_s r_s }{r_s} - \frac{ 1 + Q }{r_s} \\
        & < m_s \text{,}
    \end{align*}
    where (a) comes from the definition of $s$.
    We have
    \begin{align*}
        \ERANK(\mathcal{T}) & \OVERSET{(a)}{\geq} \ERANK(\mathcal{T}^*) \\
        & \OVERSET{(b)}{=} \sum_{j=1}^{s^*-1} m_j r_j + P r_{s^*} + Q + 1 \\
        & \OVERSET{\eqref{eq:thm:optOL:pqdef}}{=} k \text{,}
    \end{align*}
    where (a) is Lemma \ref{lem:rnkera}, and (b) is due to Lemma \ref{lem:ERANK} and Remark \ref{rem:dsum},
    hence erasure correction is possible from $\mathcal{T}$.

    The proof is complete by noting that
    substituting \eqref{eq:thm:optOL:pqdef} into \eqref{eq:thm:optOL:Tcard} yields
    \begin{equation*}
        \lvert \mathcal{T} \rvert = k + \sum_{j=1}^{s-1} m_j ( \delta_j - 1 ) + P( \delta_s - 1 ) \text{,}
    \end{equation*}
    which is equal to $\tau$ since
    \begin{equation*}
        P \OVERSET{\eqref{eq:thm:optOL:pqdef}}{=}
            \left\lfloor \frac{ k - \sum_{j=1}^{s^*-1} m_j r_j - 1 }{r_{s^*}} \right\rfloor =
            \left\lceil \frac{ k - \sum_{j=1}^{s^*-1} m_j r_j }{r_{s^*}} \right\rceil - 1 \text{.}
    \end{equation*}
\end{proof}

\begin{example} \label{ex:main:UL-ans}
    Theorem \ref{thm:UL-distbound} and \ref{thm:optOL} show that $ d \leq d_{UB} \triangleq 14 $
    is a tight upper bound on the minimum distance for the codes in Example \ref{ex:main:UL},
    and the explicit construction of Gabidulin-based LRCs with unequal locality by Construction \ref{cnstrct}
    achieves $ d = d_{UB}= 14 $.
    This is a significant improvement compared to the prior knowledge on the optimal minimum distance $d$,
    i.e., $ d_1 \triangleq 6 \leq d \leq d_2 \triangleq 16 $.
\end{example}

Note that we now also have the answer to the problem given by Example \ref{ex:UL:req} on codes with unequal $r$-locality,
which was only partially solved in Example \ref{ex:UL:prf} under the locality profile restriction.

\begin{example}
    Theorem \ref{thm:UL-distbound} and \ref{thm:optOL} show that $ d \leq d_{UB} \triangleq 8 $
    is a tight upper bound on the minimum distance for the codes in Example \ref{ex:UL:req},
    and the explicit construction of Gabidulin-based LRCs with unequal locality by Construction \ref{cnstrct}
    achieves $ d = d_{UB}= 8 $.
    This is significant improvement compared to the prior knowledge on the optimal minimum distance $d$,
    i.e., $ d_1 \triangleq 3 \leq d \leq d_2 \triangleq 9 $.
\end{example}

\section{Further Results} \label{sect:etc}

\subsection{Optimality in terms of the Bound based on Layered Locality}

In this subsection, we show that Construction \ref{cnstrct} with ordered-$(r,\delta)$ is also optimal
in terms of the distance upper bound based on \emph{layered locality} (Proposition \ref{prop:LL-distbound}).
As mentioned in the proof of Corollary \ref{cor:RBC},
the upper bounds of Theorem \ref{thm:UL-distbound} and Proposition \ref{prop:LL-distbound} coincide
under the parameter conditions of Construction \ref{cnstrct}.
Therefore, the minimum distance of codes by Construction \ref{cnstrct} that have \emph{unequal locality} with parameters
$\{(n_j,r_j,\delta_j)\}_{j\in[s^*]}$ achieve the equality of the bound by Proposition \ref{prop:LL-distbound}.
To claim optimality, we further show that the constructed codes have indeed layered locality
with the same parameters as unequal locality, after the following lemma.

\begin{lemma} \label{lem:egrank}
    For a symbol index set $ \mathcal{T} \subset [n] $ of an $[n,k]$ Gabidulin-based LRC with unequal locality
    (Construction \ref{cnstrct}), we have
    \begin{equation*}
        \GRANK(\mathcal{T}) = \min(\ERANK(\mathcal{T}),k) \text{,}
    \end{equation*}
    which implies that if either $ \GRANK(\mathcal{T}) < k $ or $ \ERANK(\mathcal{T}) < k $, then
    \begin{equation*}
        \GRANK(\mathcal{T}) = \ERANK(\mathcal{T}) \text{.}
    \end{equation*}
\end{lemma}

\begin{IEEEproof}
    See Appendix \ref{sect:appx:egrank}.
\end{IEEEproof}

\begin{proposition}
    Gabidulin-based LRCs with unequal locality satisfying the ordered $(r,\delta)$ condition
    and the condition of $ k > r_{s^*} $, are distance optimal with respect to the distance upper bound for
    codes with layered locality (Proposition \ref{prop:LL-distbound}).
\end{proposition}

\begin{proof}
    Recall that we have to show that a constructed code $\mathscr{C}$ has layered locality
    with parameters $\{(n_j,r_j,\delta_j)\}_{j\in[s^*]}$.
    Let $\mathcal{N}_j$ denote the index set of the symbols encoded by the $[r_j+\delta_j-1,r_j,\delta_j]_q$ MDS codes.
    For every $ i \in \mathcal{N}_j $, $ j \in [s^*] $, it is obvious that the $i$th symbol has $(r_j,\delta_j)$-locality
    with $\mathcal{S}_i$ being the support of the $[r_j+\delta_j-1,r_j,\delta_j]_q$ MDS local code corresponding to $i$.
    In the following, we further show by contradiction,
    that the $i$th symbol does not have $(r_{j'},\delta_{j'})$-locality for $ j' < j $.

    Suppose that there exists a symbol with index $ i \in \mathcal{N}_j $ having $(r_{j'},\delta_{j'})$-locality
    for some $ j,j' \in [s^*] $ such that $ j' < j $.
    This implies the existence of a set $\mathcal{S}_i'$ such that $ i \in \mathcal{S}_i' $, 
    $ \lvert \mathcal{S}_i' \rvert \leq r_{j'} + \delta_{j'} - 1 $,
    and $ d(\mathscr{C}\rvert_{\mathcal{S}_i'}) \geq \delta_{j'} $.
    We claim that for an arbitrary set $ \mathcal{T} \subset \mathcal{S}_i' $ such that
    $ \lvert \mathcal{T} \rvert \geq \lvert \mathcal{S}_i' \rvert - ( \delta_{j'} - 1 ) $, it must be true that
    \begin{equation}
        \ERANK(\mathcal{T}) = \ERANK(\mathcal{S}_i') \text{.} \label{eq:prop:LL-opt:1}
    \end{equation}
    First, note that $\mathcal{T}$ is an erasure correctable symbol index set for $\mathscr{C}\rvert_{S_i'}$, hence 
    $ \GRANK(\mathcal{T}) = \DIM(\mathscr{C}\rvert_{\mathcal{S}_i'}) = \GRANK(\mathcal{S}_i') $ (see Remark \ref{rem:rank}).
    Furthermore, we have
    \begin{equation*}
        \GRANK(\mathcal{S}_i') \OVERSET{(a)}{\leq} r_{j'} \OVERSET{(b)}{\leq} r_{s^*}
            \OVERSET{(c)}{<} k \text{,}
    \end{equation*}
    where (a) is by Remark \ref{rem:LCSB}, (b) is due to Definition \ref{def:ORDCON}, and (c) comes from the problem statement.
    The claim \eqref{eq:prop:LL-opt:1} follows by applying Lemma \ref{lem:egrank}.
    The remaining part of the proof proceeds with two cases.

    For the first case, assume that $ \lvert \mathcal{S}_i' \cap \mathcal{S}_i \rvert \leq \delta_{j'} - 1 $ and let
    \begin{equation*}
        \mathcal{T} = \mathcal{S}_i' \setminus \mathcal{S}_i
            = \mathcal{S}_i' \setminus ( \mathcal{S}_i' \cap \mathcal{S}_i ) \text{,}
    \end{equation*}
    so that $ \lvert \mathcal{T} \rvert = \lvert \mathcal{S}_i' \rvert - \lvert \mathcal{S}_i' \cap \mathcal{S}_i \rvert
    \geq \lvert \mathcal{S}_i' \rvert - ( \delta_{j'} - 1 ) $.
    Since $ i \in \mathcal{S}_i' $, $ \mathcal{T} \subset \mathcal{S}_i' $, and $ i \notin \mathcal{T} $,
    it follows that
    \begin{align*}
        \ERANK(\mathcal{S}_i') & \geq \ERANK(\mathcal{T} \sqcup \{i\}) \\
        & \OVERSET{(a)}{=} \ERANK(\mathcal{T}) + 1
            \text{,}
    \end{align*}
    where (a) is due to Remark \ref{rem:dsum}.
    This contradicts \eqref{eq:prop:LL-opt:1}.

    For the second case where $ \lvert \mathcal{S}_i' \cap \mathcal{S}_i \rvert \geq \delta_{j'} $,
    let $\mathcal{Q}$ and $\mathcal{Q}'$ be arbitrary sets such that
    $ \mathcal{Q},\mathcal{Q}' \subset \mathcal{S}_i' \cap \mathcal{S}_i $
    with $ \lvert \mathcal{Q} \rvert = \delta_{j'} - 1 $ and $ \lvert \mathcal{Q}' \rvert = \delta_{j'} - 2 $.
    By letting $ \mathcal{T} = \mathcal{S}_i' \setminus \mathcal{Q} $ and
    $ \mathcal{T}' = \mathcal{S}_i' \setminus \mathcal{Q}' $,
    so that $ \lvert \mathcal{T} \rvert = \lvert \mathcal{S}' \rvert - ( \delta_{j'} - 1 ) $, we get
    \begin{align*}
        \ERANK(\mathcal{T}) & = \ERANK( ( \mathcal{S}_i' \setminus \mathcal{S}_i ) \sqcup
            ( ( \mathcal{S}_i' \cap \mathcal{S}_i ) \setminus \mathcal{Q} ) )\\
        & \OVERSET{(a)}{=} \ERANK( \mathcal{S}_i' \setminus \mathcal{S}_i )
            + \ERANK( ( \mathcal{S}_i' \cap \mathcal{S}_i ) \setminus \mathcal{Q} )\\
        & \OVERSET{(b)}{=} \ERANK( \mathcal{S}_i' \setminus \mathcal{S}_i )
            + \lvert \mathcal{S}_i' \cap \mathcal{S}_i \rvert - \lvert \mathcal{Q} \rvert \text{,}
    \end{align*}
    and
    \begin{align*}
        \ERANK(\mathcal{T}') & = \ERANK( ( \mathcal{S}_i' \setminus \mathcal{S}_i ) \sqcup
            ( ( \mathcal{S}_i' \cap \mathcal{S}_i ) \setminus \mathcal{Q}' ) )\\
        & \OVERSET{(a)}{=} \ERANK( \mathcal{S}_i' \setminus \mathcal{S}_i )
            + \ERANK( ( \mathcal{S}_i' \cap \mathcal{S}_i ) \setminus \mathcal{Q}' )\\
        & \OVERSET{(c)}{=} \ERANK( \mathcal{S}_i' \setminus \mathcal{S}_i )
            + \lvert \mathcal{S}_i' \cap \mathcal{S}_i \rvert - \lvert \mathcal{Q}' \rvert \text{,}
    \end{align*}
    where (a) is due to Remark \ref{rem:dsum}, (b) follows from Lemma \ref{lem:ERANK}
    with the observation that $ ( \mathcal{S}_i' \cap \mathcal{S}_i ) \setminus \mathcal{Q} \subset \mathcal{S}_i $ and
    \begin{align*}
        \lvert ( \mathcal{S}_i' \cap \mathcal{S}_i ) \setminus \mathcal{Q} \rvert & =
            \lvert \mathcal{S}_i' \cap \mathcal{S}_i \rvert - \lvert \mathcal{Q} \rvert
            \leq \lvert \mathcal{S}_i' \rvert - \lvert \mathcal{Q} \rvert \\
        & \leq r_{j'} + \delta_{j'} - 1 - ( \delta_{j'} - 1 ) = r_{j'} \\
        & \leq r_j \text{,}
    \end{align*}
    and (c) again follows from Lemma \ref{lem:ERANK} with the following observations.
    First, consider the case where $ r_{j'} = r_{j} $.
    Due to Definition \ref{def:LL} and \ref{def:ORDCON}, we have
    \begin{equation*}
        \delta_{j'} \geq \delta_j + 1 \text{,}
    \end{equation*}
    and therefore,
    \begin{align*}
        \lvert ( \mathcal{S}_i' \cap \mathcal{S}_i ) \setminus \mathcal{Q}' \rvert & =
            \lvert \mathcal{S}_i' \cap \mathcal{S}_i \rvert - \lvert \mathcal{Q}' \rvert
            \leq \lvert \mathcal{S}_i \rvert - \lvert \mathcal{Q}' \rvert \\
        & = r_j + \delta_j - 1 - ( \delta_{j'} - 2 ) \leq r_j + \delta_j - 1 - ( \delta_j - 1 ) \\
        & = r_j \text{.}
    \end{align*}
    Otherwise, we have
    \begin{equation*}
        r_{j'} + 1 \leq r_{j} \text{,}
    \end{equation*}
    hence
    \begin{align*}
        \lvert ( \mathcal{S}_i' \cap \mathcal{S}_i ) \setminus \mathcal{Q}' \rvert & =
            \lvert \mathcal{S}_i' \cap \mathcal{S}_i \rvert - \lvert \mathcal{Q}' \rvert
            \leq \lvert \mathcal{S}_i' \rvert - \lvert \mathcal{Q}' \rvert \\
        & \leq r_{j'} + \delta_{j'} - 1 - ( \delta_{j'} - 2 ) = r_{j'} + 1 \\
        & \leq r_j \text{.}
    \end{align*}
    The proof is complete by noting that
    \begin{equation*}
        \ERANK(\mathcal{S}_i') \geq \ERANK(\mathcal{T}') > \ERANK(\mathcal{T}) \text{,}
    \end{equation*}
    which is again a contradiction to \eqref{eq:prop:LL-opt:1}.
\end{proof}

\subsection{Two Different Ordered $(r,\delta)$-locality Case}

In case where there are only two different ordered $(r,\delta)$-localities,
i.e., $\{(n_1,r_1,\delta_1),(n_2,r_2,\delta_2)\}$-locality such that $ r_1 \leq r_2 $, $ \delta_1 \geq \delta_2 $,
and $ (r_1,\delta_1) \neq (r_2,\delta_2) $, it is possible to obtain a closed form minimum distance upper bound
that is tighter than Theorem \ref{thm:UL-distbound}, given that a special condition holds.
It is given in the proposition following the corollary below,
which restates Theorem \ref{thm:UL-distbound} in a relevant form.

\begin{corollary} \label{cor:TDL}
    \begin{sloppypar}
    The minimum Hamming distance of codes having unequal locality with parameters $\{(n_1,r_1,\delta_1),(n_2,r_2,\delta_2)\}$
    satisfying the ordered $(r,\delta)$ condition is upper bounded by
    \begin{enumerate}
        \item if $ \lfloor m_1 \rfloor r_1 \geq k $,
            \begin{equation*}
                d \leq n - k + 1 - \left( \left\lceil \frac{k}{r_1} \right\rceil - 1 \right)
                    ( \delta_1 - 1 ) \text{,}
            \end{equation*}
        \item if $ \lfloor m_1 \rfloor r_1 < k $,
            \begin{equation*}
                d \leq n - k + 1 - \lfloor m_1 \rfloor ( \delta_1 - 1 )
                    - \left( \left\lceil \frac{ k - \lfloor m_1 \rfloor r_1}{r_2} \right\rceil - 1 \right) ( \delta_2 - 1 )
                    \text{.}
            \end{equation*}
    \end{enumerate}
    \end{sloppypar}
\end{corollary}

\begin{proposition} \label{prop:TDL}
    \begin{sloppypar}
    The minimum Hamming distance of codes having unequal locality with parameters $\{(n_1,r_1,\delta_1),(n_2,r_2,\delta_2)\}$
    satisfying the ordered $(r,\delta)$ condition and also a special condition such that
    $ q_1 = 0 $ or $ \delta_1 - 1 \leq q_1 \leq r_1 + \delta_1 - 2 $, is upper bounded by
    \begin{enumerate}
        \item if $ \lceil m_1 \rceil r_1 \geq k $,
            \begin{equation}
                d \leq n - k + 1 - \left( \left\lceil \frac{k}{r_1} \right\rceil - 1 \right)
                    ( \delta_1 - 1 ) \text{,} \label{eq:prop:TDL:b1}
            \end{equation}
        \item if $ \lceil m_1 \rceil r_1 < k $,
            \begin{equation}
                d \leq n - k + 1 - \lceil m_1 \rceil ( \delta_1 - 1 )
                    - \left( \left\lceil \frac{ k - \lceil m_1 \rceil r_1 }{r_2} \right\rceil - 1 \right) ( \delta_2 - 1 )
                    \text{.} \label{eq:prop:TDL:b2}
            \end{equation}
    \end{enumerate}
    \end{sloppypar}
\end{proposition}

\begin{proof}
    Recall that the codes also have layered locality with parameters $\{(\hat{n}_1,r_1,\delta_1),(\hat{n}_2,r_2,\delta_2)\}$.
    By Proposition \ref{prop:LL-distbound}, we therefore have
    \begin{enumerate}
        \item if $ \hat{k}_1 \geq k $,
            \begin{equation}
                d \leq n - k + 1 - \left( \left\lceil \frac{k}{r_1} \right\rceil - 1 \right)
                    ( \delta_1 - 1 ) \text{,} \label{eq:prop:TDL:p1}
            \end{equation}
        \item if $ \hat{k}_1 < k $,
            \begin{equation}
                d \leq n - k + 1 - ( \hat{n}_1 - \hat{k}_1 ) 
                    - \left( \left\lceil \frac{ k - \hat{k}_1 }{r_2} \right\rceil - 1 \right) ( \delta_2 - 1 ) \text{.}
                    \label{eq:prop:TDL:p2}
            \end{equation}
    \end{enumerate}
    First, consider the case where $ \hat{k} \geq k $, hence \eqref{eq:prop:TDL:p1} holds.
    Since the bounds by \eqref{eq:prop:TDL:p1} and \eqref{eq:prop:TDL:b1} are identical,
    we only have to check that \eqref{eq:prop:TDL:p1} implies \eqref{eq:prop:TDL:b2},
    given that $ \lceil m_1 \rceil r_1 < k $.
    This can be easily seen by noting that
    \begin{align*}
        \left( \left\lceil \frac{k}{r_1} \right\rceil - 1 \right)( \delta_1 - 1 ) & = \lceil m_1 \rceil ( \delta_1 - 1 ) +
            \left( \left\lceil \frac{ k - \lceil m_1 \rceil r_1 }{r_1} \right\rceil - 1 \right) ( \delta_1 - 1 ) \\
        & \geq \lceil m_1 \rceil ( \delta_1 - 1 ) +
            \left( \left\lceil \frac{ k - \lceil m_1 \rceil r_1 }{r_2} \right\rceil - 1 \right) ( \delta_2 - 1 ) \text{.}
    \end{align*}
    We thus have to further verify the claim for the second case of $ \hat{k}_1 < k $.

    \emph{Case 1}: $ \lceil m_1 \rceil r_1 \geq k $.\\
    If $ 0 \leq \hat{q}_1 \leq \delta_1 - 2 $, then
    \begin{align}
        \hat{n}_1 - \hat{k}_1 & = \hat{n}_1 - \lfloor \hat{m}_1 \rfloor r_1 \geq \hat{n}_1 - \hat{m}_1 r_1
            = \hat{m}_1 ( \delta_1 - 1 ) \notag \\
        & \OVERSET{(a)}{\geq} \lceil m_1 \rceil ( \delta_1 - 1 ) \text{,} \label{eq:prop:TDL:nmrlb}
    \end{align}
    where (a) is due to Lemma \ref{lem:layered-n:cor} and the conditions imposed on $\hat{q}_1$ and $q_1$.

    If otherwise $ \delta_1 - 1 \leq \hat{q}_1 \leq r_1 + \delta_1 - 2 $, again we have \eqref{eq:prop:TDL:nmrlb} as
    \begin{align*}
        \hat{n}_1 - \hat{k}_1 & = \lceil \hat{m}_1 \rceil ( \delta_1 - 1 ) \\
        & \OVERSET{(a)}{\geq} \lceil m_1 \rceil ( \delta_1 - 1 ) \text{,}
    \end{align*}
    where (a) is again due to Lemma \ref{lem:layered-n:cor}.

    Furthermore, from the condition $ \lceil m_1 \rceil r_1 \geq k $, we have
    \begin{equation}
        \lceil m_1 \rceil \geq \frac{k}{r_1} > \left\lceil \frac{k}{r_1} \right\rceil - 1 \text{.}
            \label{eq:prop:TDL:cmj1lb}
    \end{equation}
    Substituting \eqref{eq:prop:TDL:nmrlb} with \eqref{eq:prop:TDL:cmj1lb} into \eqref{eq:prop:TDL:p2},
    and removing the last subtrahend that is non-negative, results in \eqref{eq:prop:TDL:b1}.

    \emph{Case 2}: $ \lceil m_1 \rceil r_1 < k $.\\
    If $ 0 \leq \hat{q}_1 \leq \delta_1 - 2 $, we have
    \begin{align*}
        \hat{n}_1 - \hat{k}_1 & = \hat{n}_1 - \lfloor \hat{m}_1 \rfloor r_1 \geq \hat{n}_1 - \hat{m}_1 r_1
            = \hat{m}_1 ( \delta_1 - 1 ) \\
        & \geq \lfloor \hat{m}_1 \rfloor ( \delta_1 - 1 ) \text{.}
    \end{align*}
    Substituting into \eqref{eq:prop:TDL:p2}, and also considering the definition of $\hat{k}_1$, we get
    \begin{equation*}
        d \leq n - k + 1 - \lfloor \hat{m}_1 \rfloor ( \delta_1 - 1 ) -
            \left( \left\lceil \frac{ k - \lfloor \hat{m}_1 \rfloor r_1 }{r_2} \right\rceil - 1 \right)
            ( \delta_2 - 1 ) \text{.}
    \end{equation*}
    One can verify that, for integers $a,b$ such that $ a \geq b $, we have
    \begin{equation}
        a( \delta_1 - 1 ) + \left\lceil \frac{ k - a r_1 }{r_2} \right\rceil ( \delta_2 - 1 ) \geq
            b( \delta_1 - 1 ) + \left\lceil \frac{ k - b r_1 }{r_2} \right\rceil ( \delta_2 - 1 ) \text{,}
            \label{eq:prop:TDL:ab}
    \end{equation}
    due to the ordered $(r,\delta)$ condition.
    Therefore, noting that $ \lfloor \hat{m}_1 \rfloor \geq \lceil m_1 \rceil $ for the same reason as in
    \eqref{eq:prop:TDL:nmrlb}, we get \eqref{eq:prop:TDL:b2}.

    Otherwise, if $ \delta_1 - 1 \leq \hat{q}_1 \leq r_1 + \delta_1 - 2 $, we have
    \begin{align*}
        \hat{k}_1 & = \hat{n}_1 - \lceil \hat{m}_1 \rceil ( \delta_1 - 1 ) \leq \hat{n}_1 - \hat{m}_1 ( \delta - 1 )
            = \hat{m}_1 j_1 \\
        & \leq \lceil \hat{m}_1 \rceil j_1 \text{.}
    \end{align*}
    Again, substituting into \eqref{eq:prop:TDL:p2}, and also considering the definition of $\hat{k}_1$, we get
    \begin{equation*}
        d \leq n - k + 1 - \lceil \hat{m}_1 \rceil ( \delta_1 - 1 ) -
            \left( \left\lceil \frac{ k - \lceil \hat{m}_1 \rceil r_1 }{r_2} \right\rceil - 1 \right) ( \delta_2 - 1 )
            \text{,}
    \end{equation*}
    and therefore \eqref{eq:prop:TDL:b2} due to \eqref{eq:prop:TDL:ab} and Lemma \ref{lem:layered-n:cor}.
\end{proof}

It is easy to verify that the upper bound of Proposition \ref{prop:TDL} implies the bound by Corollary \ref{cor:TDL}.
Furthermore, the following example shows the existence of cases where the bound by Proposition \ref{prop:TDL}
is strictly tighter, hence is a tighter bound.
However, in the parameter regime of our optimal code construction (Construction \ref{cnstrct}),
Proposition \ref{prop:TDL} coincides with Corollary \ref{cor:TDL}, and also with
the minimum distance of the optimal code construction.

\begin{example} \label{ex:TDL}
    Consider codes having unequal locality with parameters $\{(n_1=5,r_1=2,\delta_1=2),(n_2=10,r_2=3,\delta_2=2)\}$.
    The minimum distance upper bound of $ d \leq 9 $ given by Proposition \ref{prop:TDL} is strictly tighter than
    that of Corollary \ref{cor:TDL} which is $ d \leq 10 $.
\end{example}

For the special case of $ \delta_1 = \delta_2 = 2 $, the condition of Proposition \ref{prop:TDL} such that
$ q_1 = 0 $ or $ \delta_1 - 1 \leq q_1 \leq r_1 + \delta_1 - 2 $, is always true (as in Example \ref{ex:TDL})
and therefore can be omitted.
In this case, the bound by Proposition \ref{prop:TDL} becomes identical to \cite[Thm. 6]{Zeh16ISIT}.%
\footnote{
    There is a slight deviation in the boundary conditions,
    but one can check that this makes no difference and both bounds coincide.
}
However, note that \cite[Thm. 6]{Zeh16ISIT} is more restrictive (for $ r_1 < r_2 $) in that \emph{disjointness} is assumed.

\section{Conclusion} \label{sect:conclusion}

In this work, we have investigated the minimum distance characteristics of LRCs with unequal $(r,\delta)$-locality.
The problem has been analyzed in both the layered locality and ordinary (non-layered) locality scenario.
Singleton-type minimum distance bounds have been presented and their tightness has been shown by an optimal construction
achieving the equality in the bounds.
Feasible rate regions have also been characterized by the dimension upper bounds that do not depend on the minimum distance.

\appendices

\section{Proof of Lemma \ref{lem:rnkera}} \label{sect:appx:rnkera}

\begin{algorithm} [t]
    \caption{Used in the Proof of Lemma \ref{lem:rnkera}} \label{alg:R}
    \begin{algorithmic}[1]
        \WHILE{ $ \exists l_1,l_2 \in [\lvert \mathcal{L} \rvert] $, $ l_1 < l_2 $, such that
            $ \lvert \mathcal{R} \cap \mathcal{G}_{l_1} \rvert < \lvert \mathcal{G}_{l_1} \rvert $ and
            $ \lvert \mathcal{R} \cap \mathcal{G}_{l_2} \rvert > 0 $ }
            \STATE Construct $\Delta\mathcal{R}_1$ and $\Delta\mathcal{R}_2$ such that \\
                $ \Delta\mathcal{R}_1 \subset \mathcal{G}_{l_1} \setminus \mathcal{R} $, \\
                $ \Delta\mathcal{R}_2 \subset \mathcal{R} \cap \mathcal{G}_{l_2} $, and \\
                $ \lvert \Delta\mathcal{R}_1 \rvert = \lvert \Delta\mathcal{R}_2 \rvert =
                    \min( \lvert \mathcal{G}_{l_1} \setminus \mathcal{R} \rvert ,
                    \lvert \mathcal{R} \cap \mathcal{G}_{l_2} \rvert ) $
                \STATE $ \mathcal{R} = \mathcal{R} \sqcup \Delta\mathcal{R}_1 \setminus \Delta\mathcal{R}_2 $
                \label{alg:R:step}
        \ENDWHILE
    \end{algorithmic}
\end{algorithm}

\begin{proof}
    Any erasure pattern can be transformed into the claimed worst case pattern by repeatedly invoking Algorithm \ref{alg:R}.
    This is because, in Step \ref{alg:R:step} of the algorithm,
    symbols as many as possible in the local group $\mathcal{G}_{l_2}$
    are replaced with symbols in the local group $\mathcal{G}_{l_1}$, where $ l_1 < l_2 $.
    We show that that this replacement always results in a non-increasing \emph{remaining rank},
    making the claimed pattern worst indeed.

    First, observe that
    \begin{equation*}
        \mathcal{R} = \mathcal{R}_0 \sqcup \mathcal{R}_1 \sqcup \mathcal{R}_2 \text{,}
    \end{equation*}
    where
    \begin{align*}
        \mathcal{R}_0 & = \mathcal{R} \setminus ( \mathcal{R}_1 \sqcup \mathcal{R}_2 ) \text{,} \\
        \mathcal{R}_1 & = \mathcal{R} \cap \mathcal{G}_{l_1} \text{,} \\
        \mathcal{R}_2 & = \mathcal{R} \cap \mathcal{G}_{l_2} \text{.}
    \end{align*}
    Due to Remark \ref{rem:dsum}, we have
    \begin{equation}
        \ERANK(\mathcal{R}) = \ERANK( \mathcal{R}_0 ) + \ERANK( \mathcal{R}_1 ) + \ERANK( \mathcal{R}_2 ) \text{.}
            \label{eq:lem:rnkera:r}
    \end{equation}
    Similarly, for
    \begin{align*}
        \mathcal{R}' & \triangleq \mathcal{R} \sqcup \Delta\mathcal{R}_1 \setminus \Delta\mathcal{R}_2 \\
        & = \mathcal{R}_0 \sqcup \mathcal{R}_1' \sqcup \mathcal{R}_2' \text{,}
    \end{align*}
    where
    \begin{align*}
        \mathcal{R}_1' & = \mathcal{R}' \cap \mathcal{G}_{l_1} = \mathcal{R}_1 \sqcup \Delta\mathcal{R}_1 \text{,} \\
        \mathcal{R}_2' & = \mathcal{R}' \cap \mathcal{G}_{l_2} = \mathcal{R}_2 \setminus \Delta\mathcal{R}_2 \text{,}
    \end{align*}
    we can write
    \begin{equation}
        \ERANK(\mathcal{R}') = \ERANK( \mathcal{R}_0 ) + \ERANK( \mathcal{R}_1' ) + \ERANK( \mathcal{R}_2' ) \text{.}
            \label{eq:lem:rnkera:rp}
    \end{equation}
    From \eqref{eq:lem:rnkera:r} and \eqref{eq:lem:rnkera:rp}, we have to show that
    \begin{equation}
        \ERANK(\mathcal{R}_1) + \ERANK(\mathcal{R}_2) \geq \ERANK(\mathcal{R}_1') + \ERANK({\mathcal{R}_2'}) \text{.}
            \label{eq:lem:rnkera}
    \end{equation}

    Let $\mathcal{G}_{l_i}$ be of $(r_{j_i},\delta_{j_i})$, $ i = 1,2 $.
    By the ordering of $\mathcal{L}$ and the ordered $(r,\delta)$ condition, we have
    $ r_{j_1} \leq r_{j_2} $ and $ \delta_{j_1} \geq \delta_{j_2} $.
    Note that by Lemma \ref{lem:ERANK}, we have
    \begin{align}
        \begin{split}
            \ERANK(\mathcal{R}_1) & = \min( \lvert \mathcal{R}_1 \rvert, r_{j_1} ) \text{,} \\
            \ERANK(\mathcal{R}_2) & = \min( \lvert \mathcal{R}_2 \rvert, r_{j_2} ) \text{,}
        \end{split} \label{eq:lem:rnkera:rnkr} \\
        \begin{split}
            \ERANK(\mathcal{R}_1') & = \min( \lvert \mathcal{R}_1 \rvert + \Delta, r_{j_1} ) \text{,} \\
            \ERANK(\mathcal{R}_2') & = \min( \lvert \mathcal{R}_2 \rvert - \Delta, r_{j_2} ) \text{,}
        \end{split} \label{eq:lem:rnkera:rnkrp}
    \end{align}
    where
    \begin{align}
        \Delta & = \lvert \Delta\mathcal{R}_1 \rvert = \lvert \Delta\mathcal{R}_2 \rvert \notag \\
        & = \min( \lvert \mathcal{G}_{l_1} \setminus \mathcal{R} \rvert , \lvert \mathcal{R} \cap \mathcal{G}_{l_2} \rvert )
            \notag \\
        & = \min( \lvert \mathcal{G}_{l_1} \rvert - \lvert \mathcal{R}_1 \rvert, \lvert \mathcal{R}_2 \rvert )
            \text{.} \label{eq:lem:rnkera:D}
    \end{align}
    We only provide the proof for the case where $ \lvert \mathcal{R}_1 \rvert \leq r_{j_1} $ and
    $ \lvert \mathcal{R}_2 \rvert > r_{j_2} $,
    since it is easy to verify that \eqref{eq:lem:rnkera} holds in other cases.
    From \eqref{eq:lem:rnkera:rnkr}, we have
    \begin{equation*}
        \ERANK(\mathcal{R}_1) + \ERANK(\mathcal{R}_2) = \lvert \mathcal{R}_1 \rvert + r_{j_2} \text{.}
    \end{equation*}
    If $ \lvert \mathcal{G}_{l_2} \rvert - \lvert \mathcal{R}_1 \rvert \geq \lvert \mathcal{R}_2 \rvert $,
    from \eqref{eq:lem:rnkera:D} and \eqref{eq:lem:rnkera:rnkrp}, we get
    \begin{equation*}
        \ERANK(\mathcal{R}_1') + \ERANK(\mathcal{R}_2') \leq r_{j_1} \leq r_{j_2}
            \leq \lvert \mathcal{R}_1 \rvert + r_{j_2} \text{,}
    \end{equation*}
    and therefore \eqref{eq:lem:rnkera}.
    Otherwise, \eqref{eq:lem:rnkera} again holds since
    \begin{align*}
        \ERANK(\mathcal{R}_1') + \ERANK(\mathcal{R}_2') & \leq
            r_{j_1} + \lvert \mathcal{R}_1 \rvert + \lvert \mathcal{R}_2 \rvert - \lvert \mathcal{G}_{l_1} \rvert \\
        & = \lvert \mathcal{R}_1 \rvert + \lvert \mathcal{R}_2 \rvert - ( \delta_{j_1} - 1 ) \\
        & \leq \lvert \mathcal{R}_1 \rvert + \lvert \mathcal{G}_{l_2} \rvert - ( \delta_{j_2} - 1 ) \\
        & = \lvert \mathcal{R}_1 \rvert + r_{j_2} \text{.}
    \end{align*}
\end{proof}

\section{Proof of Lemma \ref{lem:egrank}} \label{sect:appx:egrank}

\begin{proof}
    \begin{sloppypar}
    Let us denote the evaluation points corresponding to $\mathcal{T}$ as $\{y_1,\ldots,y_{\lvert \mathcal{T} \rvert}\}$,
    where $ y_i \in \mathbb{F}_q^t $ (or equivalently $ y_i \in \mathbb{F}_{q^t} $),
    $ i \in [ \lvert \mathcal{T} \rvert ] $.
    Without loss of generality, we assume that the set $\{ y_1, \ldots, y_{\ERANK(\mathcal{T})} \}$
    is a basis for the vector space $\SPAN(\{y_1,\ldots,y_{\lvert \mathcal{T} \rvert}\})$.
    Then, for $ i = \ERANK(\mathcal{T}) + 1, \ldots, \lvert \mathcal{T} \rvert $, we have
    \begin{equation*}
        y_i = \sum_{j=1}^{\ERANK(\mathcal{T})} \lambda_{ij} y_j \text{,}
    \end{equation*}
    where $ \lambda_{ij} \in \mathbb {F}_q $.
    The generator submatrix corresponding to the symbols indexed by $\mathcal{T}$ can be written as
    \begin{equation*}
        G\rvert_\mathcal{T} \triangleq ( \mathbf{g}_1^T\ \ldots\ \mathbf{g}_{\lvert \mathcal{T} \rvert}^T ) = \begin{pmatrix}
            y_1^{q^0}     & \cdots & y_{\lvert \mathcal{T} \rvert}^{q^0}\\
            \vdots                 & \ddots & \vdots\\
            y_1^{q^{k-1}} & \cdots & y_{\lvert \mathcal{T} \rvert}^{q^{k-1}}
        \end{pmatrix}\text{.}
    \end{equation*}
    Furthermore, for $i = \ERANK(\mathcal{T}) + 1 , \ldots , \lvert \mathcal{T} \rvert $, we can write
    \begin{equation*}
        \mathbf{g}_i^T =
            \begin{pmatrix}
                ( \sum_j \lambda_{ij} y_j )^{q^0} \\
                \vdots \\
                ( \sum_j \lambda_{ij} y_j )^{q^{k-1}}
            \end{pmatrix}
        \OVERSET{\eqref{eq:fqlin}}{=}
            \begin{pmatrix}
                \sum_j \lambda_{ij} y_j^{q^0} \\
                \vdots \\
                \sum_j \lambda_{ij} y_j^{q^{k-1}}
            \end{pmatrix}
        = \sum_{j=1}^{\ERANK(\mathcal{T})} \lambda_{ij} \mathbf{g}_j^T \text{,}
    \end{equation*}
    and therefore
    \begin{equation*}
        \GRANK(\mathcal{T}) = \GRANK(\mathcal{T}') = \RANK(G\rvert_{\mathcal{T}'}) \text{,}
    \end{equation*}
    where $\mathcal{T}'$ is the symbol index set corresponding to the evaluation points
    $\{ y_1, \ldots, y_{\ERANK(\mathcal{T})} \}$, i.e.,
    $ G\rvert_{\mathcal{T}'} = ( \mathbf{g}_1^T \ \ldots \ \mathbf{g}_{\ERANK(\mathcal{T})}^T ) $.
    Note that $G\rvert_{\mathcal{T}'}$ is a \emph{Moore matrix} \cite{Goss98Book,Macwilliams77Book} of size
    $k \times \ERANK(\mathcal{T})$ with all the elements in the first row being linearly independent over $\mathbb{F}_q$.
    The proof is complete by considering the fact that any arbitrary square submatrix of $ G\rvert_{\mathcal{T}'} $
    is nonsingular.
    \end{sloppypar}
\end{proof}

\bibliographystyle{IEEEtran}
\bibliography{IEEEabrv,final}

\begin{thebibliography}{10}
\providecommand{\url}[1]{#1}
\csname url@samestyle\endcsname
\providecommand{\newblock}{\relax}
\providecommand{\bibinfo}[2]{#2}
\providecommand{\BIBentrySTDinterwordspacing}{\spaceskip=0pt\relax}
\providecommand{\BIBentryALTinterwordstretchfactor}{4}
\providecommand{\BIBentryALTinterwordspacing}{\spaceskip=\fontdimen2\font plus
\BIBentryALTinterwordstretchfactor\fontdimen3\font minus
  \fontdimen4\font\relax}
\providecommand{\BIBforeignlanguage}[2]{{%
\expandafter\ifx\csname l@#1\endcsname\relax
\typeout{** WARNING: IEEEtran.bst: No hyphenation pattern has been}%
\typeout{** loaded for the language `#1'. Using the pattern for}%
\typeout{** the default language instead.}%
\else
\language=\csname l@#1\endcsname
\fi
#2}}
\providecommand{\BIBdecl}{\relax}
\BIBdecl

\bibitem{Kim16ALC}
G.~Kim and J.~Lee, ``Local erasure correction codes with unequal locality
  profile,'' in \emph{2016 54th Annual Allerton Conference on Communication,
  Control, and Computing (Allerton)}, Sept 2016.

\bibitem{Sathiamoorthy13ICVLDB}
M.~Sathiamoorthy, M.~Asteris, D.~Papailiopoulos, A.~G. Dimakis, R.~Vadali,
  S.~Chen, and D.~Borthakur, ``Xoring elephants: novel erasure codes for big
  data,'' in \emph{Proceedings of the 39th international conference on Very
  Large Data Bases}, 2013, pp. 325--336.

\bibitem{Gopalan12TIT}
P.~Gopalan, C.~Huang, H.~Simitci, and S.~Yekhanin, ``On the locality of
  codeword symbols,'' \emph{IEEE Transactions on Information Theory}, vol.~58,
  no.~11, pp. 6925--6934, Nov 2012.

\bibitem{Prakash12ISIT}
N.~Prakash, G.~M. Kamath, V.~Lalitha, and P.~V. Kumar, ``Optimal linear codes
  with a local-error-correction property,'' in \emph{Information Theory
  Proceedings (ISIT), 2012 IEEE International Symposium on}, July 2012, pp.
  2776--2780.

\bibitem{Kamath14TIT}
G.~M. Kamath, N.~Prakash, V.~Lalitha, and P.~V. Kumar, ``Codes with local
  regeneration and erasure correction,'' \emph{IEEE Transactions on Information
  Theory}, vol.~60, no.~8, pp. 4637--4660, Aug 2014.

\bibitem{Papailiopoulos14TIT}
D.~S. Papailiopoulos and A.~G. Dimakis, ``Locally repairable codes,''
  \emph{IEEE Transactions on Information Theory}, vol.~60, no.~10, pp.
  5843--5855, Oct 2014.

\bibitem{Tamo14TIT}
I.~Tamo and A.~Barg, ``A family of optimal locally recoverable codes,''
  \emph{IEEE Transactions on Information Theory}, vol.~60, no.~8, pp.
  4661--4676, Aug 2014.

\bibitem{Kadhe16ISIT}
S.~Kadhe and A.~Sprintson, ``Codes with unequal locality,'' in \emph{2016 IEEE
  International Symposium on Information Theory (ISIT)}, July 2016, pp.
  435--439.

\bibitem{Kadhe16ARX}
\BIBentryALTinterwordspacing
------, ``Codes with unequal locality,'' \emph{CoRR}, vol. abs/1601.06153,
  2016. [Online]. Available: \url{http://arxiv.org/abs/1601.06153}
\BIBentrySTDinterwordspacing

\bibitem{Zeh16ISIT}
A.~Zeh and E.~Yaakobi, ``Bounds and constructions of codes with multiple
  localities,'' in \emph{2016 IEEE International Symposium on Information
  Theory (ISIT)}, July 2016, pp. 640--644.

\bibitem{Zeh16ARX}
\BIBentryALTinterwordspacing
------, ``Bounds and constructions of codes with multiple localities,''
  \emph{CoRR}, vol. abs/1601.02763, 2016. [Online]. Available:
  \url{http://arxiv.org/abs/1601.02763}
\BIBentrySTDinterwordspacing

\bibitem{Kuijper14ARX}
\BIBentryALTinterwordspacing
M.~Kuijper and D.~Napp, ``Erasure codes with simplex locality,'' \emph{CoRR},
  vol. abs/1403.2779, 2014. [Online]. Available:
  \url{http://arxiv.org/abs/1403.2779}
\BIBentrySTDinterwordspacing

\bibitem{Tamo13ISIT}
I.~Tamo, D.~S. Papailiopoulos, and A.~G. Dimakis, ``Optimal locally repairable
  codes and connections to matroid theory,'' in \emph{Information Theory
  Proceedings (ISIT), 2013 IEEE International Symposium on}, July 2013, pp.
  1814--1818.

\bibitem{Silberstein13ISIT}
N.~Silberstein, A.~S. Rawat, O.~O. Koyluoglu, and S.~Vishwanath, ``Optimal
  locally repairable codes via rank-metric codes,'' in \emph{Information Theory
  Proceedings (ISIT), 2013 IEEE International Symposium on}, July 2013, pp.
  1819--1823.

\bibitem{Song14JSAC}
W.~Song, S.~H. Dau, C.~Yuen, and T.~J. Li, ``Optimal locally repairable linear
  codes,'' \emph{IEEE Journal on Selected Areas in Communications}, vol.~32,
  no.~5, pp. 1019--1036, May 2014.

\bibitem{Goparaju14ISIT}
S.~Goparaju and R.~Calderbank, ``Binary cyclic codes that are locally
  repairable,'' in \emph{2014 IEEE International Symposium on Information
  Theory}, June 2014, pp. 676--680.

\bibitem{Tamo15ISIT}
I.~Tamo, A.~Barg, S.~Goparaju, and R.~Calderbank, ``Cyclic lrc codes and their
  subfield subcodes,'' in \emph{2015 IEEE International Symposium on
  Information Theory (ISIT)}, June 2015, pp. 1262--1266.

\bibitem{Hao16ISIT}
J.~Hao, S.~T. Xia, and B.~Chen, ``Some results on optimal locally repairable
  codes,'' in \emph{2016 IEEE International Symposium on Information Theory
  (ISIT)}, July 2016, pp. 440--444.

\bibitem{Ernvall16TIT}
T.~Ernvall, T.~Westerback, R.~Freij-Hollanti, and C.~Hollanti, ``Constructions
  and properties of linear locally repairable codes,'' \emph{IEEE Transactions
  on Information Theory}, vol.~62, no.~3, pp. 1129--1143, March 2016.

\bibitem{Poellaenen16ISIT}
A.~Pollanen, T.~Westerback, R.~Freij-Hollanti, and C.~Hollanti, ``Bounds on the
  maximal minimum distance of linear locally repairable codes,'' in \emph{2016
  IEEE International Symposium on Information Theory (ISIT)}, July 2016, pp.
  1586--1590.

\bibitem{Chen16ARX}
\BIBentryALTinterwordspacing
B.~Chen, S.-T. Xia, J.~Hao, and F.-W. Fu, ``Constructions of optimal cyclic
  $(r,\delta)$ locally repairable codes,'' \emph{CoRR}, vol. abs/1609.01136,
  2016. [Online]. Available: \url{http://arxiv.org/abs/1609.01136}
\BIBentrySTDinterwordspacing

\bibitem{Song15ARX}
\BIBentryALTinterwordspacing
W.~Song and C.~Yuen, ``Locally repairable codes with functional repair and
  multiple erasure tolerance,'' \emph{CoRR}, vol. abs/1507.02796, 2015.
  [Online]. Available: \url{http://arxiv.org/abs/1507.02796}
\BIBentrySTDinterwordspacing

\bibitem{Gabidulin85}
{\`E}.~M. Gabidulin, ``Theory of codes with maximum rank distance,''
  \emph{Problemy Peredachi Informatsii}, vol.~21, no.~1, pp. 3--16, 1985.

\bibitem{Macwilliams77Book}
F.~MacWilliams and N.~Sloane, \emph{The Theory of Error Correcting
  Codes}.\hskip 1em plus 0.5em minus 0.4em\relax North-Holland Publishing
  Company, 1977.

\bibitem{Rawat14TIT}
A.~S. Rawat, O.~O. Koyluoglu, N.~Silberstein, and S.~Vishwanath, ``Optimal
  locally repairable and secure codes for distributed storage systems,''
  \emph{IEEE Transactions on Information Theory}, vol.~60, no.~1, pp. 212--236,
  Jan 2014.

\bibitem{Chen17ISIT}
B.~Chen, S.~T. Xia, and J.~Hao, ``Locally repairable codes with multiple
  $(r_i,\delta_i)$-localities,'' in \emph{2017 IEEE International Symposium on
  Information Theory (ISIT)}, June 2017, pp. 2038--2042.

\bibitem{Goss98Book}
D.~Goss, \emph{Basic structures of function field arithmetic}.\hskip 1em plus
  0.5em minus 0.4em\relax Berlin New York: Springer, 1998.

\end{thebibliography}

\end{document}